\RequirePackage{fix-cm}
\documentclass[smallextended]{svjour3}       
\smartqed  
\usepackage{graphicx}
\usepackage{cite}
\usepackage{amsmath,amssymb,amsfonts}
\usepackage{algorithmic}
\usepackage{textcomp}
\usepackage{xcolor}
\usepackage{subcaption}
\usepackage{multirow}
\usepackage{algorithm}
\usepackage{comment}
\usepackage{url}
\usepackage{natbib}
\newtheorem{mydef}{Definition}
\begin{document}

\title{\textit{struc2gauss}: Structural Role Preserving Network Embedding via Gaussian Embedding}


\author{Yulong Pei         \and
        Xin Du \and
        Jianpeng Zhang \and \\
        George Fletcher \and
        Mykola Pechenizkiy
}


\institute{Y. Pei\and X. Du \and G. Fletcher\and M. Pechenizkiy \at
              Department of Mathematics and Computer Science \\
              Eindhoven University of Technology, 5600 MB Eindhoven, the Netherlands\\
              \email{\{y.pei.1,x.du,g.h.l.fletcher,m.pechenizkiy\}@tue.nl}           
            \and
            J. Zhang \at
            National Digital Switching System Engineering  Technology R\&D Center\\
            Zhengzhou, China\\
            \email{zjp@ndsc.com.cn}
}

\date{Received: date / Accepted: date}

\maketitle

\begin{abstract}
Network embedding (NE) is playing a principal role in network mining, due to its ability to map nodes into efficient low-dimensional embedding vectors. However, two major limitations exist in state-of-the-art NE methods: \textbf{role preservation} and \textbf{uncertainty modeling}. Almost all previous methods represent a node into a point in space and focus on local structural information, i.e., neighborhood information. However, neighborhood information does not capture global structural information and point vector representation fails in modeling the uncertainty of node representations. In this paper, we propose a new NE framework, \textit{struc2gauss}, which learns node representations in the space of Gaussian distributions and performs network embedding based on global structural information. \textit{struc2gauss} first employs a given node similarity metric to measure the global structural information, then generates structural context for nodes and finally learns node representations via Gaussian embedding. Different structural similarity measures of networks and energy functions of Gaussian embedding are investigated. Experiments conducted on real-world networks demonstrate that \textit{struc2gauss} effectively captures global structural information while state-of-the-art network embedding methods fail to, outperforms other methods on the structure-based clustering and classification task and provides more information on uncertainties of node representations.
\keywords{Gaussian Embedding \and Structural Similarity \and Uncertainty Modeling}
\end{abstract}

\section{Introduction}
\label{intro}
Network analysis consists of numerous tasks including community detection~\citep{fortunato2010community}, role discovery~\citep{rossi2015role}, link prediction~\citep{liben2007link}, etc. As relations exist between nodes that disobey the \textit{i.i.d} assumption, it is non-trivial to apply traditional data mining techniques in networks directly. Network embedding (NE) fills the gap by mapping nodes in a network into a low-dimensional space according to their structural information in the network. It has been reported that using embedded node representations can achieve promising performance on many network analysis tasks~\citep{perozzi2014deepwalk,grover2016node2vec,cao2015grarep,ribeiro2017struc2vec}.

Previous NE techniques mainly relied on eigendecomposition~\citep{shaw2009structure,tenenbaum2000global}, but the high computational complexity of eigendecomposition makes it difficult to apply in real-world networks. With the fast development of neural network techniques, unsupervised embedding algorithms have been widely used in natural language processing (NLP) where words or phrases from the vocabulary are mapped to vectors in the learned embedding space, e.g., \textit{word2vec}~\citep{mikolov2013efficient,mikolov2013distributed} and GloVe~\citep{pennington2014glove}. By drawing an analogy between paths consists of several nodes on networks and word sequences in text, DeepWalk~\citep{perozzi2014deepwalk} learns node representations based on random walks using the same mechanism of \textit{word2vec}. Afterwards, a sequence of studies have been conducted to improve DeepWalk either by extending the definition of neighborhood to higher-order proximity~\citep{cao2015grarep,tang2015line,grover2016node2vec,perozzi2016walklets} or incorporating more information for node representations such as attributes~\citep{li2017attributed,wang2017attributed} and heterogeneity~\citep{chang2015heterogeneous,tang2015pte}.

\begin{figure}
\centering
\includegraphics[width=2.8in]{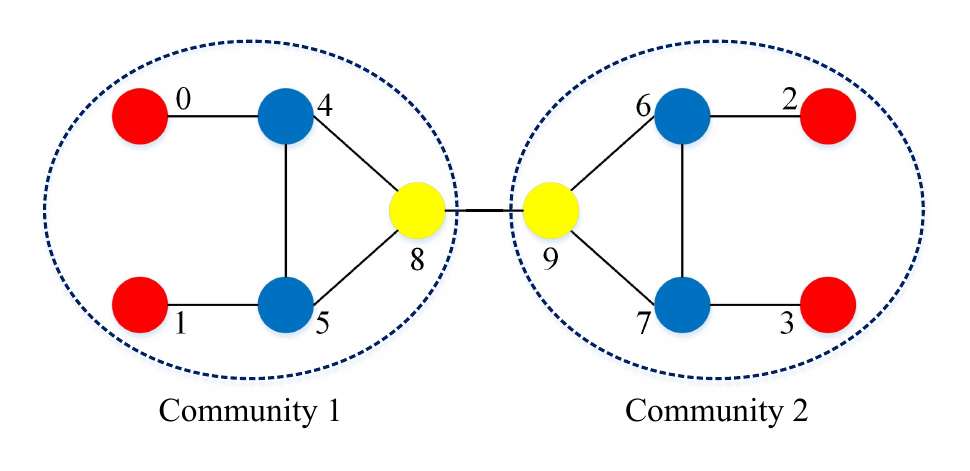}
\caption{An example of ten nodes belonging to (1) three groups (different colors indicate different groups) based on global structural information, i.e., the structural roles and (2) two groups (groups are shown by the dashed ellipses) based on local structural information, i.e., the communities. For example, nodes 0, 1, 4, 5 and 8 belong to the same group Community 1 based on local structural perspective because they have more internal connections. Node 0 and 2 are far from each other, but they are in the same group based on global structural perspective.}
\label{fig:exp}
\end{figure}

Although a variety of NE methods have been proposed, two major limitations exist in previous NE studies: \textbf{role preservation} and \textbf{uncertainty modeling}. Previous methods focused only on one of these two limitations and while neglecting the other. In particular, for \textbf{role preservation}, most studies applied random walk to learn representations. However, random walk based embedding strategies and their higher-order extensions can only capture local structural information, i.e., first-order and higher-order proximity within the neighborhood of the target node~\citep{lyu2017enhancing}. Local structural information is reflected in community structures of networks. But these methods may fail in capturing global structural information, i.e., structural roles~\citep{rossi2015role,pei2018dynmf}. Global structural information represents roles of nodes in networks, where two nodes have the same role if they are structurally similar from a global perspective. An example of global structural information (roles) and local structural information (communities) is shown in Fig.~\ref{fig:exp}. In summary, nodes that belong to the same community require dense local connections while nodes that have the same role may have no common neighbors at all~\citep{tu2018deep}. Empirical evidence based on this example for illustrating this limitation will be shown in Section~\ref{case}. For \textbf{uncertainty modeling}, most previous methods represented a node into a point vector in the learned embedding space. However, real-world networks may be noisy and imbalanced. For example, node degree distributions in real-world networks are often skewed where some low-degree nodes may contain less discriminative information~\citep{tu2018deep}. Point vector representations learned by these methods are deterministic~\citep{dos2016multilabel} and are not capable of modeling the uncertainties of node representations.

There are a few studies trying to address these limitations in the literature. For instance, \textit{struc2vec}~\citep{ribeiro2017struc2vec} builds a hierarchy to measure similarity at different scales, and constructs a multilayer graph to encode the structural similarities. \textit{SNS}~\citep{lyu2017enhancing} discovers graphlets as a pre-processing step to obtain the structural similar nodes. DRNE~\citep{tu2018deep} learns network embedding by modeling regular equivalence~\citep{wasserman1994social}. However, these studies aim only to solve the problem of \textbf{role preservation} to some extent. Thus the limitation of \textbf{uncertainty modeling} remains a challenge. \citep{dos2016multilabel} and \citep{bojchevski2017deep} put effort in improving classification tasks by embedding nodes into Gaussian distributions but both methods only capture the neighborhood information based on random walk techniques. DVNE~\citep{zhu2018deep} learns Gaussian embedding for nodes in the Wasserstein space as the latent representations to capture uncertainties of nodes, but they focus only on first- and second-order proximity of networks same to previous methods. Therefore, the problem of \textbf{role preservation} has not been solved in these studies.

In this paper, we propose \textit{struc2gauss}, a new structural role preserving network embedding framework. \textit{struc2gauss} learns node representations in the space of Gaussian distributions and performs NE based on global structural information so that it can address both limitations simultaneously. On the one hand, \textit{struc2gauss} generates node context based on a global structural similarity measure to learn node representations so that global structural information can be taken into consideration. On the other hand, \textit{struc2gauss} learns node representations via Gaussian embedding and each node is represented as a Gaussian distribution where the mean indicates the position of this node in the embedding space and the covariance represents its uncertainty. Furthermore, we analyze and compare two different energy functions for Gaussian embedding to calculate the closeness of two embedded Gaussian distributions, i.e., expected likelihood and KL divergence. To investigate the influence of structural information, we also compare \textit{struc2gauss} to two other structural similarity measures for networks, i.e., MatchSim and SimRank.

We summarize the contributions of this paper as follows:
\begin{itemize}
    \item We propose a flexible structure preserving network embedding framework, \textit{struc2gauss}, which learns node representations in the space of Gaussian distributions. \textit{struc2gauss} is capable of preserving structural roles and modeling uncertainties.
    \item We investigate the influence of different energy functions in Gaussian embedding and compare to different structural similarity measures in preserving global structural information of networks.
    \item We conduct extensive experiments in node clustering and classification tasks which demonstrate the effectiveness of \textit{struc2gauss} in capturing the global structural role information of networks and modeling the uncertainty of learned node representations.
\end{itemize}

The rest of the paper is organized as follows. Section~\ref{related} provides an overview of the related work. We present the problem statement in Section~\ref{notations}. Section~\ref{s2g} explains the technical details of \textit{struc2gauss}. In Section \ref{exp} we then discuss our experimental study. The possible extensions of \textit{struc2gauss} are discussed in Section \ref{dis}. Finally, in Section \ref{conc} we draw conclusions and outline directions for future work.

\section{Related Work}
\label{related}

\subsection{Network Embedding}
Network embedding methods map nodes in a network into a low-dimensional space according to their structural information in the network. The learned node representations can boost performance in many network analysis tasks, e.g., community detection and link prediction. Previous methods mainly viewed NE as part of dimensionality reduction techniques~\citep{goyal2018graph}. They first construct a pairwise similarity graph based on neighborhood and then embed the nodes of the graph into a  lower dimensional vector space. Locally Linear Embedding (LLE)~\citep{tenenbaum2000global} and Laplacian Eigenmaps~\citep{belkin2002laplacian} are two representative methods in this category. SPE~\citep{shaw2009structure} learns a low-rank kernel matrix to capture the structures of input graph via a set of linear inequalities as constraints. But the high computational complexity makes these methods difficult to apply in real-world networks.

With increasing attention attracted by neural network research, unsupervised neural network techniques have opened up a new world for embedding. \textit{word2vec} as well as Skip-Gram and CBOW~\citep{mikolov2013efficient,mikolov2013distributed} learn low-rank representations of words in text based on word context and show promising results of different NLP tasks. Based on \textit{word2vec}, DeepWalk~\citep{perozzi2014deepwalk} first introduces such embedding mechanism to networks by treating nodes as words and random walks as sentences.  Afterwards, a sequence of studies have been conducted to improve DeepWalk either by extending the definition of neighborhood to higher-order proximity~\citep{cao2015grarep,tang2015line,grover2016node2vec,perozzi2016walklets} or incorporating more information for node representations such as attributes~\citep{li2017attributed,wang2017attributed} and heterogeneity~\citep{chang2015heterogeneous,tang2015pte}. Recently, deeper neural networks have also been introduced in NE problem to capture the non-linear characteristics of networks, such as SDNE~\citep{wang2016structural}. However, these approaches represent a node into a point vector in the learned embedding space and are not capable of modeling the uncertainties of node representations. To solve this problem, inspired by~\citep{vilnis2014word}, Gaussian embedding has been used in NE. \citep{bojchevski2017deep} learns node embeddings by leveraging Gaussian embedding to capture uncertainties. \citep{dos2016multilabel} combines Gaussian embedding and classification loss function for multi-label network classification. DVNE~\citep{zhu2018deep} learns a Gaussian embedding for each node in the Wasserstein space as the latent representation so that the uncertainties can be modeled. We refer the reader to~\citep{hamilton2017repre,cui2018survey,cai2018comprehensive} for more details.

\begin{small}
\begin{table}
\centering
\caption{A brief summary of different NE methods. Note that (1) we only list NE methods for homogeneous networks without attributes, and (2) \textit{node2vec}~\citep{grover2016node2vec} aims to capture both local and global structure information but walk-based sampling strategy does not effectively capture global structure information, as shown in our experiments in Section~\ref{exp}.}
\label{tb:summary}
\begin{tabular}{|l|c|c|c|c|}
\hline
Method                   & \begin{tabular}[c]{@{}c@{}}community\\ (local)\end{tabular} & \begin{tabular}[c]{@{}c@{}}role\\ (global)\end{tabular}  & uncertainty \\ \hline
DeepWalk~\citep{perozzi2014deepwalk}         &   $\surd$         &                          &             \\ \hline
LINE~\citep{tang2015line} &     $\surd$             &               &                        \\ \hline
GraRep~\citep{cao2015grarep}       &   $\surd$             &                          &             \\ \hline
PTE~\citep{tang2015pte}       &    $\surd$           &                     &             \\ \hline
Walklets~\citep{perozzi2016walklets}         &   $\surd$         &                          &             \\ \hline
\textit{node2vec}~\citep{grover2016node2vec}    &  $\surd$           &                     &             \\ \hline
EP~\citep{duran2017learning}    &  $\surd$           &                     &             \\ \hline
GraphSage~\citep{hamilton2017inductive}    &  $\surd$           &                     &             \\ \hline
\textit{struc2vec}~\citep{ribeiro2017struc2vec} &       &     $\surd$             &             \\ \hline
DRNE~\citep{tu2018deep}                &                   &   $\surd$             &             \\ \hline
GraphWave~\citep{donnat2018learning}                &                   &   $\surd$             &             \\ \hline
DVNE~\citep{zhu2018deep}               &    $\surd$            &                         &  $\surd$            \\ \hline
SNS~\citep{lyu2017enhancing}           &                   &    $\surd$               &             \\ \hline
our method &                   &     $\surd$          &    $\surd$          \\ \hline
\end{tabular}
\end{table}
\end{small}

Recent years have witnessed increasing interest in neural networks on graphs. Graph neural networks~\citep{scarselli2008graph} can also learn node representations but using more complicated operations such as convolution. \citep{kipf2016semi} proposes a GCN model using an efficient layer-wise propagation rule based on a first-order approximation of spectral convolutions on graphs. \citep{gilmer2017neural} introduces a general message passing neural network framework to interpret different previous neural models for graphs. GraphSAGE~\citep{hamilton2017inductive} learns node representations in an inductive manner sampling a fixed-size neighborhood of each node, and then performing a specific aggregator over it. Embedding Propagation (EP)~\citep{duran2017learning} learns representations of graphs by passing messages forward and backward in an unsupervised setting. Graph Attention Networks (GATs)~\citep{velickovic2017graph} extend graph convolutions by utilizing masked self-attention layers to assign different importances to different nodes with different sized neighborhoods.

However, most NE methods as well as graph neural networks only concern the local structural information represented by paths consists of linked nodes, i.e., the community structures of networks. But they fail to capture global structural information, i.e., structural roles. SNS~\citep{lyu2017enhancing}, \textit{struc2vec}~\citep{ribeiro2017struc2vec} and DRNE~\citep{tu2018deep} are exceptions which take global structural information into consideration. SNS uses graphlet information for structural similarity calculation as a pre-propcessing step. \textit{struc2vec} applies the dynamic time warping to measure similarity between two nodes' degree sequences and builds a new multilayer graph based on the similarity. Then similar mechanism used in DeepWalk has been used to learn node representations. DRNE explicitly models regular equivalence, which is one way to define the structural role, and leverages the layer normalized LSTM~\citep{ba2016layer} to learn the representations for nodes. Another related work focusing on global structural information is REGAL~\citep{heimann2018regal}. REGAL aims at matching nodes across different graphs so the global structural patterns should be considered. However, its target is network alignment but not representation learning. A brief summary of these NE methods is list in Table~\ref{tb:summary}.

\subsection{Structural Similarity}
\label{strucsim}
Structure based network analysis tasks can be categorized into two types: structural similarity calculation and network clustering . 

Calculating structural similarities between nodes is a hot topic in recent years and different methods have been proposed. SimRank~\citep{jeh2002simrank} is one of the most representative notions to calculate structural similarity. It implements a recursive definition of node similarity based on the assumption that two objects are similar if they relate to similar objects. SimRank++~\citep{antonellis2008simrank++} adds an 
evidence weight which partially compensates for the neighbor matching cardinality problem. P-Rank~\citep{zhao2009p} extends SimRank by jointly encoding both in- and out-link relationships into structural similarity computation. MatchSim~\citep{lin2009matchsim} uses maximal matching of neighbors to calculate the structural similarity. RoleSim~\citep{jin2011axiomatic} is the only similarity measure which can satisfy the automorphic equivalence properties.

Network clusters can be based on either global or local structural information. Graph clustering based on global structural information is the problem of role discovery~\citep{rossi2015role}. In social science research, roles are represented as concepts of equivalence~\citep{wasserman1994social}. Graph-based methods and feature-based methods have been proposed for this task. Graph-based methods take nodes and edges as input and directly partition nodes into groups based on their structural patterns. For example, Mixed Membership Stochastic Blockmodel~\citep{airoldi2008mixed} infers the role distribution of each node using the Bayesian generative model. Feature-based methods first transfer the original network into feature vectors and then use clustering methods to group nodes. For example, RolX~\citep{henderson2012rolx} employs ReFeX~\citep{henderson2011s} to extract features of networks and then uses non-negative matrix factorization to cluster nodes. Local structural information based clustering corresponds to the problem of community detection~\citep{fortunato2010community}. A community is a group of nodes that interact with each other more frequently than with those outside the group. Thus, it captures only local connections between nodes.

\section{Problem Statement}
\label{notations}
We illustrated local community structure and global role structure in Section~\ref{intro} using the example in Fig.~\ref{fig:exp}. In this section, definitions of community and role will be presented and then we formally define the problem of structural role preserving network embedding.

Structural role is from social science and used to describe nodes in a network from a global perspective. Formally,
\begin{mydef}
\label{roledef}
\textbf{Structural role}. In a network, a set of nodes have the same role if they share similar structural properties
(such as degree, clustering coefficient, and betweenness) and structural roles can often be associated with various functions in a network.
\end{mydef}
For example, hub nodes with high degree in a social network are more likely to be opinion leaders, whereas bridge nodes with high betweenness are gatekeepers to connect different groups. Structural roles can reflect the global structural information because two nodes which have the same role could be far from each other and have no direct links or shared neighbors. In contrast to roles, community structures focus on local connections between nodes.
\begin{mydef}
\textbf{Community structure}. In a network, communities can represent the local structures of nodes, i.e., the organization of nodes in communities, with many edges joining nodes of the same community and comparatively few edges joining nodes of different communities~\citep{fortunato2010community}. A community is a set of nodes where nodes in this set are densely connected internally.
\end{mydef}
It can be seen that the focus of community structure is the internal and local connections so it aims to capture the local structural information of networks

In this study, we only consider the global structural information, i.e., structural role information, so without mentioning it explicitly, structural information indicates the global one and the keyphrases ``structural role information" and ``global structural information" are used interchangeably.

\begin{mydef}
\textbf{Structural Role Preserving Network Embedding}. Given a network $G = (V, E)$, where $V$ is a set of nodes and $E$ is a set of edges between the nodes, the problem of \textbf{Structural Preserving Network Embedding} aims to represent each node $v\in V$ into a Gaussian distribution with mean $\mu$ and covariance $\Sigma$ in a low-dimensional space $\mathbb{R}^d$, i.e., learning a function 
$$f: V\to \mathcal{N}(x;\mu,\Sigma),$$
where $\mu\in \mathbb{R}^d$ is the mean, $\Sigma\in\mathbb{R}^{d\times d}$ is the covariance and $d\ll |V|$. In the space $\mathbb{R}^d$, the global structural role information of nodes introduced in Definition~\ref{roledef} can be preserved, i.e., if two nodes have the same role their means should be similar, and the uncertainty of node representations can be captured, i.e., the values of variances indicate the levels of uncertainties of learned representations.
\end{mydef}

\section{\textit{struc2gauss}}
\begin{figure*}
\centering
\includegraphics[width=4.7in]{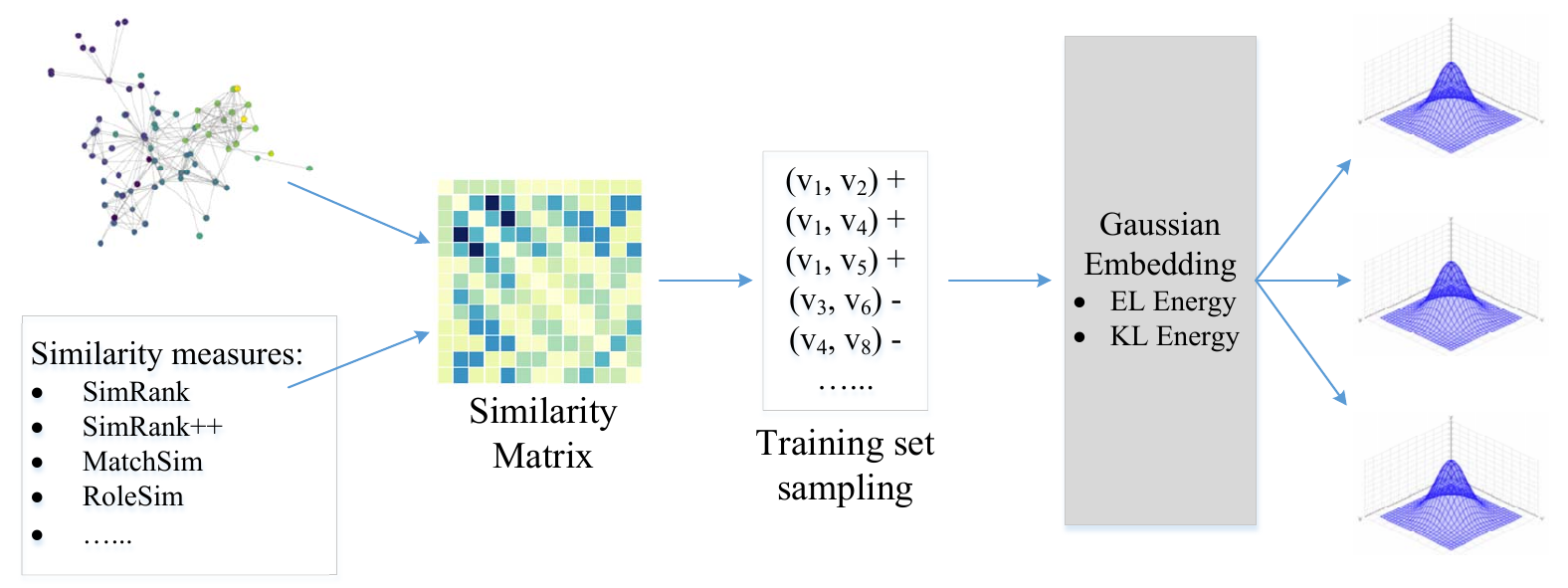}
\caption{Overview of the \textit{struc2gauss} framework. \textit{struc2gauss} consists of three components: similarity calculation, training set sampling and Gaussian embedding.}
\label{fig:frame}
\end{figure*}

An overview of our proposed \textit{struc2gauss} framework is shown in Fig.~\ref{fig:frame}. Given a network, a similarity measure is employed to calculate the similarity matrix, then the training set which consists of positive and negative pairs are sampled based on the similarity matrix. Finally, Gaussian embedding techniques are applied on the training set and generate the embedded Gaussian distributions as the node representations and uncertainties of the representations. Besides, we analyze the computational complexity and the flexibility of our \textit{struc2gauss} framework.
\label{s2g}

\subsection{Structural Similarity Calculation}
\label{simmatrix}
It has been theoretically proved that random walk sampling based NE methods are not capable of capturing structural equivalence~\citep{lyu2017enhancing} which is one way to model the structural roles in networks~\citep{wasserman1994social}. Thus, to capture the global structural information, we calculate the pairwise structural similarity as a pre-processing step similar to \citep{lyu2017enhancing,ribeiro2017struc2vec}. 

In the literature, a variety of structural similarity measures have been proposed to calculate node similarity based on the structures of networks, e.g., SimRank~\citep{jeh2002simrank}, MatchSim~\citep{lin2009matchsim} and RoleSim~\citep{jin2011axiomatic,jin2014scalable}. However, not all of these measures can capture the global structural role information and we will show the empirical evidence in the experiments in Section~\ref{exp}. Therefore, in this paper we leverage RoleSim for the structural similarity since it satisfies all the requirements of Axiomatic Role Similarity Properties for modeling the equivalence~\citep{jin2011axiomatic}, i.e., the structural roles. RoleSim also generalizes Jaccard coefficient and corresponds linearly to the maximal weighted matching. RoleSim similarity between two nodes $u$ and $v$ is defined as:
\begin{align}
    &RoleSim(u,v)=(1-\beta)\max_{M(u,v)}\frac{\sum_{(x,y)\in M(u,v)}RoleSim(x,y)}{|N(u)|+|N(v)|-|M(u,v)|}+\beta
\end{align}
where $|N(u)|$ and $|N(v)|$ are the numbers of neighbors of node $u$ and $v$, respectively. $M(u,v)$ is a matching between
$N(u)$ and $N(v)$, i.e., $M(u, v)\subseteq N(u)\times N(v)$ is a bijection between $N(u)$ and $N(v)$. The parameter $\beta$ is a decay factor where $0 < \beta < 1$. The intuition of RoleSim is that two nodes are structurally similar if their corresponding neighbors are also structurally similar. This intuition is consistent with the notion of automorphic and regular equivalence~\citep{wasserman1994social}.

In practice, RoleSim values can be computed iteratively and are guaranteed to converge. The procedure of computing RoleSim consists of three steps:
\begin{itemize}
    \item Step 1: Initialize matrix of RoleSim scores $R^0$;
    \item Step 2: Compute the $k^{th}$ iteration $R^k$ scores for the $(k-1)^{th}$ iteration's values, $R^{k-1}$ using:
    \begin{align}
        &R^{k}(u,v)=(1-\beta)\max_{M(u,v)}\frac{\sum_{(x,y)\in M(u,v)}R^{k-1}(x,y)}{|N(u)|+|N(v)|-|M(u,v)|}+\beta
    \end{align}
    \item Step 3: Repeat Step 2 until $R$ values converge for each pair of nodes.
\end{itemize}
Note that there are other strategies can be used to capture the global structural role information except structural similarity, and these possible strategies will be discussed in Section~\ref{dis}. The advantage of RoleSim in capturing structural roles to other structural measures will also be discussed empirically in Section~\ref{sim}.

\subsection{Training Set Sampling}
The target of structural role preserving network embedding is to map nodes in the network to a latent space where the learned latent representations of two nodes are (1) more similar if these two nodes are structurally similar, and (2) more dissimilar if these two nodes are not structurally similar. Hence, we need to generate structurally similar and dissimilar node pairs as the training set based on the similarity we learned in Section~\ref{simmatrix}. We name the structurally similar pairs of nodes the positive set and the structurally dissimilar pairs the negative set.

In detail, for node $v$, we rank its similarity values towards other nodes and then select top-$k$ most similar nodes $u_i,i=1,...,k$ to form its positive set $\Gamma_{+}=\{(v,u_i)|i=1,...,k\}$. For the negative set, we randomly select the same number of nodes $\{u'_i,i=1,...,k\}$ same to~\citep{vilnis2014word} and other random walk sampling based methods~\citep{perozzi2014deepwalk,tang2015line,grover2016node2vec}, i.e., $\Gamma_{-}=\{(v,u'_i)|i=1,...,k\}$. Therefore, $k$ is a parameter indicating the \textit{number of positive/negative nodes per node}. We will generate $r$ positive and negative sets for each node where $r$ is a parameter indicating the \textit{number of samples per node}. The influence of these parameters will be analyzed empirically in Section~\ref{paramstuning}. Note that the selection of the positive set is similar to that in DeepWalk and the difference is that we follow the similarity rank to select the positive nodes instead of random walks.

\subsection{Gaussian Embedding}
\label{gaussemb}

\subsubsection{Overview}
\label{over}
Recently language modeling techniques such as \textit{word2vec} have been extensively used to learn word representations in and almost all NE studies are based on these word embedding techniques. However, these NE studies map each entity to a fixed point vector in a low-dimension space so that the uncertainties of learned embeddings are ignored. Gaussian embedding aims to solve this problem by learning density-based distributed embeddings in the space of Gaussian distributions~\citep{vilnis2014word}. Gaussian embedding has been utilized in different graph mining tasks including triplet classification on knowledge graphs~\citep{he2015learning}, multi-label classification on heterogeneous graphs~\citep{dos2016multilabel} and link prediction and node classification on attributed graphs~\citep{bojchevski2017deep}.

Gaussian embedding trains with a ranking-based loss based on the ranks of positive and negative samples. Following~\citep{vilnis2014word}, we choose the max-margin ranking objective which can push scores of positive pairs above negatives by a margin defined as:
\begin{equation}
    \mathcal{L}=\sum_{(v,u)\in \Gamma_{+}}\sum_{(v',u')\in \Gamma_{-}}\max(0, m-\mathcal{E}(z_v, z_u)+\mathcal{E}(z_{v'}, z_{u'}))
\end{equation}
where $\Gamma_{+}$ and $\Gamma_{-}$ are the positive and negative pairs, respectively. $\mathcal{E}(\cdot,\cdot)$ is the energy function which is used to measure the similarity of two distributions, $z_v$ and $z_u$ are the learned Gaussian distributions for nodes $v$ and $u$, and $m$ is the margin separating positive and negative pairs. In this paper, we present two different energy functions to measure the similarity of two distributions for node representation learning, i.e., expected likelihood and KL divergence based energy functions. For the learned Gaussian distribution $z_i\sim\mathcal{N}(0;\mu_i,\Sigma_i)$ for node $i$, to reduce the computational complexity, we restrict the covariance matrix $\Sigma_i$ to be diagonal and spherical in this work.

\subsubsection{Expected Likelihood based Energy}

Although both dot product and inner product can be used to measure similarity between two distributions, dot product only considers means and does not incorporate covariances. Thus, we use inner product to measure the similarity. Formally, the integral of inner product between two Gaussian distributions $z_i$ and $z_j$ (learned Gaussian embeddings for node $i$ and $j$ respectively), a.k.a., expected likelihood, is defined as:
\begin{align}
    \label{eq:el}
    E(z_i,z_j)&=\int_{x\in \mathbb{R}}\mathcal{N}(x;\mu_i,\Sigma_i)\mathcal{N}(x;\mu_j,\Sigma_j)dx=\mathcal{N}(0;\mu_i-\mu_j,\Sigma_i+\Sigma_j).
\end{align}
For simplicity in computation and comparison, we use the logarithm of Eq.~(\ref{eq:el}) as the final energy function:
\begin{align}
    \label{eq:logel}
    &\mathcal{E}_{EL}(z_i,z_j)=\log E(z_i,z_j)=\log \mathcal{N}(0;\mu_i-\mu_j,\Sigma_i+\Sigma_j)\\\nonumber
    =&\frac{1}{2}\Big\{(\mu_i-\mu_j)^T(\Sigma_i+\Sigma_j)^{-1}(\mu_i-\mu_j)+\log\det(\Sigma_i+\Sigma_j)+d\log(2\pi)\Big\}
\end{align}
where $d$ is the number of dimensions. The gradient of this energy function with respect to the means $\mu$ and covariances $\Sigma$ can be calculated in a closed form as:
\begin{align}
\label{el}
    \frac{\partial\mathcal{E}_{EL}(z_i,z_j)}{\partial\mu_i}&=-\frac{\partial\mathcal{E}(z_i.z_j)_{EL}}{\partial\mu_j}=-\Delta_{ij}\\\nonumber
    \frac{\partial\mathcal{E}_{EL}(z_i,z_j)}{\partial\Sigma_i}&=\frac{\partial\mathcal{E}(z_i.z_j)_{EL}}{\partial\Sigma_j}=\frac{1}{2}(\Delta_{ij}\Delta_{ij}^T-(\Sigma_i+\Sigma_j)^{-1}),
\end{align}
where $\Delta_{ij}=(\Sigma_i+\Sigma_j)^{-1}(\mu_i-\mu_j)$~\citep{he2015learning,vilnis2014word}.
Note that expected likelihood is a symmetric similarity measure, i.e., $\mathcal{E}_{EL}(z_i,z_j)=\mathcal{E}_{EL}(z_j,z_i)$.

\subsubsection{KL Divergence based Energy}

KL divergence is another straightforward way to measure the similarity between two distributions so we utilize the energy function $\mathcal{E}_{KL}(z_i,z_j)$ based on the KL divergence to measure the similarity between Gaussian distributions $z_i$ and $z_j$ (learned Gaussian embeddings for node $i$ and $j$ respectively):
\begin{align}
    &\mathcal{E}_{KL}(z_i,z_j)=D_{KL}(z_i,z_j)\\\nonumber
    =&\int_{x\in \mathbb{R}}\mathcal{N}(x;\mu_i,\Sigma_i)\log\frac{\mathcal{N}(x;\mu_j,\Sigma_j)}{\mathcal{N}(x;\mu_i,\Sigma_i)}dx\\\nonumber
    =&\frac{1}{2}\Big\{tr(\Sigma_i^{-1}\Sigma_j)+(\mu_i-\mu_j)^T\Sigma_i^{-1}(\mu_i-\mu_j)-\log\frac{\det(\Sigma_j)}{\det(\Sigma_i)}-d\Big\}
\end{align}
where $d$ is the number of dimensions. Similarly, we can compute the gradients of this energy function with respect to the means $\mu$ and covariances $\Sigma$:
\begin{align}
\label{kl}
    \frac{\partial\mathcal{E}_{KL}(z_i,z_j)}{\partial\mu_i}&=-\frac{\partial\mathcal{E}_{KL}(z_i.z_j)}{\partial\mu_j}=-\Delta_{ij}^{\prime}\\\nonumber
    \frac{\partial\mathcal{E}_{KL}(z_i,z_j)}{\partial\Sigma_i}&=\frac{1}{2}(\Sigma_i^{-1}\Sigma_j\Sigma_i^{-1}+\Delta_{ij}^{\prime}\Delta_{ij}^{\prime T}-\Sigma_i^{-1})\\\nonumber
    \frac{\partial\mathcal{E}_{KL}(z_i,z_j)}{\partial\Sigma_j}&=\frac{1}{2}(\Sigma_j^{-1}-\Sigma_i^{-1})
\end{align}
where $\Delta_{ij}^{\prime}=\Sigma_i^{-1}(\mu_i-\mu_j)$.

Note that KL divergence based energy is asymmetric but we can easily extend to a symmetric similarity measure as follows:
\begin{equation}
    \mathcal{E}(z_i,z_j)=\frac{1}{2}(D_{KL}(z_i,z_j)+D_{KL}(z_j,z_i)).
\end{equation}

\subsection{Learning}

To avoid the means to grow too large and ensure the covariances to be positive definite as well as reasonably sized, we regularize the means and covariances to learn the embedding~\citep{vilnis2014word}. Due to the different geometric characteristics, two different hard constraint strategies have been used for means and covariances, respectively. Note that we only consider diagonal and spherical covariances. In particular, we have 
\begin{equation}
\label{mean}
    \|\mu_i\|\leq C,~\forall i
\end{equation}
\begin{equation}
\label{covar}
    c_{min}I\prec \Sigma_i \prec c_{max}I,~\forall i.
\end{equation}
The constraint on means guarantees them to be sufficiently small and constraint on covariances ensures that they are positive definite and of appropriate size. For example, $\Sigma_{ii}\gets \max(c_{min},\min(c_{max},\Sigma_{ii}))$ can be used to regularize diagonal covariances.

We use AdaGrad~\citep{duchi2011adaptive} to optimize the parameters. The learning procedure is described in Algorithm~\ref{alg}. Initialization phase is from line 1 to 4, context generation is shown in line 7, and Gaussian embeddings are learned from line 8 to 14.
\renewcommand{\algorithmicrequire}{\textbf{Input:}}
\renewcommand{\algorithmicensure}{\textbf{Output:}}
\begin{algorithm}
\caption{The Learning Algorithm of \textit{struc2gauss}}
\begin{algorithmic}[1]
\label{alg}
    \REQUIRE An energy function $\mathcal{E}(z_i,z_j)$, a graph $G=(V,E)$, embedding dimension $d$, constraint values $C$ for mean and $c_{max}$ and $c_{min}$ for covariance, learning rate $\alpha$, and maximum epochs $n$.
    \ENSURE Gaussian embeddings (mean vector $\mu$ and covariance matrix $\Sigma$) for nodes $v\in V$
    \FORALL{$v\in V$}
    \STATE Initialize mean $\mu$ for $v$
    \STATE Initialize covariance $\Sigma$ for $v$
    \STATE Regularize $\mu$ and $\Sigma$ with constraint in Eq.~(\ref{mean}) and (\ref{covar})
    \ENDFOR
    \WHILE{not reach the maximum epochs $n$}
    \STATE Generate positive and negative sets $\Gamma_{+}$ and $\Gamma_{-}$ for each node
    \IF{use expected likelihood based energy}
    \STATE Update means and covariances based on Eq.~(\ref{el})
    \ENDIF
    \IF{use KL divergence based energy}
    \STATE Update means and covariances based on Eq.~(\ref{kl})
    \ENDIF
    \STATE Regularize $\mu$ and $\Sigma$ with constraint in Eq.~(\ref{mean}) and (\ref{covar})
    \ENDWHILE
\end{algorithmic}
\end{algorithm}

\subsection{Computational Complexity}
\label{complex}
The complexity of different components of \textit{struc2gauss} are analyzed as follows:
\begin{itemize}
    \item[1] For structural similarity calculation using RoleSim, the computational complexity is $O(kn^2d)$, where $n$ is the number of nodes, $k$ is the number of iterations and $d$ is the average of $y\log y$ over all node-pair bipartite graph in $G$~\citep{jin2011axiomatic} where $y=|N(u)|\times |N(v)|$ for each pair of nodes $u$ and $v$. The complexity $O(y\log y)$ is from the complexity of the fast greedy algorithm offers a $\frac{1}{2}$-approximation of the globally optimal matching.
    \item[2] To generate the training set based on similarity matrix, we need to sample from the most similar nodes for each node, i.e., to select $k$ largest numbers from an unsorted array. Using heap, the complexity is $O(n\log k)$.
    \item[3] For Gaussian embedding, the operations include matrix addition, multiplication and inversion. In practice, as stated above, we only consider two types of covariance matrices, i.e., diagonal and spherical, so all these operations have the complexity of $O(n)$.
\end{itemize}
Overall, the component of similarity calculation is the bottleneck of the framework. One possible and effective way to optimize this part is to set the similarity to be 0 if two nodes have a large difference in degrees. The reason is: (1) we generate the context only based on most similar nodes; and (2) two nodes are less likely to be structural similar if their degrees are very different.

\section{Experiments}
\label{exp}
We evaluate \textit{struc2gauss} in different tasks in order to understand its effectiveness in capturing structural information, capability in modeling uncertainties of embeddings and stability of the model towards parameters. We also study the influence of different similarity measures empirically. The source code of \textit{struc2gauss} is available online\footnote{\url{https://bitbucket.org/paulpei/struc2gauss/src/master/}}.
 
\subsection{Experimental Setup}
\subsubsection{Datasets}
We conduct experiments on two types of network datasets: networks with and without ground-truth labels where these labels can represent the global structural role information of nodes in the networks. For networks with labels, to compare to state-of-the-art, we use air-traffic networks from~\citep{ribeiro2017struc2vec} where the networks are undirected, nodes are airports, edges indicate the existence of commercial flights and labels correspond to their levels of activities. For networks without labels, we select five real-world networks in different domains from Network Repository\footnote{\url{http://networkrepository.com/index.php}}. A brief introduction to these datasets is shown in Table~\ref{tb:data}. Note that the numbers of groups for networks without labels are determined by Minimum Description Length (MDL)~\citep{henderson2012rolx}.

\begin{table}
\small
\centering
\caption{A brief introduction to data sets.}
\label{tb:data}
\begin{tabular}{|l|l|c|c|c|}
\hline
Type                                                                      & Dataset       & \# nodes & \# edges & \# groups \\ \hline
\multirow{3}{*}{\begin{tabular}[c]{@{}l@{}}with\\ labels\end{tabular}}    & Brazilian-air & 131      & 1038     & 4         \\ \cline{2-5}
                                                                          & European-air  & 399      & 5995     & 4         \\ \cline{2-5}
                                                                          & USA-air       & 1190     & 13599    & 4         \\ \hline
\multirow{5}{*}{\begin{tabular}[c]{@{}l@{}}without\\ labels\end{tabular}} & Arxiv GR-QC   & 5242     & 28980    & 8         \\ \cline{2-5}
                                                                          & Advogato      & 6551     & 51332    & 11         \\ \cline{2-5}
                                                                          & Hamsterster   & 2426     & 16630    & 10         \\ \cline{2-5}
                                                                          & Anybeat   & 12645     & 67053    & 15         \\ \cline{2-5}
                                                                          & Epinion   & 26588     & 100120    & 18         \\ \hline
\end{tabular}
\end{table}

\subsubsection{Baselines}
We compare \textit{struc2gauss} with several state-of-the-art NE methods.
\begin{itemize}
    \item DeepWalk~\citep{perozzi2014deepwalk}:  DeepWalk~\citep{perozzi2014deepwalk} learns node representations based on random walks using the same mechanism of \textit{word2vec} by drawing an analogy between paths consists of several nodes on networks and word sequences in text. The structural information is captured by the paths of nodes generated by random walks.
    \item \textit{node2vec}~\citep{grover2016node2vec}: It extends DeepWalk to learn latent representations from the node paths generated by biased random walk. Two hyper-parameters $p$ and $q$ are used to control the random walk to be breadth-first or depth-first. In this way, \textit{node2vec} can capture the structural information in networks. Note that when $p=q=1$, \textit{node2vec} degrades to DeepWalk.
    \item LINE~\citep{tang2015line}: It learns node embeddings via preserving both the local and global network structures. By extending DeepWalk, LINE aims to capture both the first-order, i.e., the neighbors of nodes, and second-order proximities, i.e., the shared neighborhood structures of nodes.
    \item Embedding Propagation (EP)~\citep{duran2017learning}: EP is an unsupervised learning framework for network embedding and learns vector representations of graphs by passing two types of messages between neighboring nodes. EP, as one of graph neural networks, is similar to graph convolutional networks (GCN)~\citep{kipf2016semi}. The difference is that EP is unsupervised and GCN is designed for semi-supervised learning.
    \item \textit{struc2vec}~\citep{ribeiro2017struc2vec}: It learns latent representations for the structural identity of nodes. Due to its high computational complexity, we use the combination of all optimizations proposed in the paper for large networks.
    \item \textit{graph2gauss}~\citep{vilnis2014word}: It maps each node into a Gaussian distribution where the mean indicates the position of a node in the embedded space and the covariance denotes the uncertainty of the learned representation. \citep{bojchevski2017deep} and~\citep{dos2016multilabel} extend the original Gaussian embedding method to network embedding task.
    \item DRNE~\citep{tu2018deep}: It learns node representations based on the concept of regular equivalence. DRNE utilizes a layer normalized LSTM to represent each node by aggregating the representations of their neighborhoods in a recursive way so that the global structural information can be preserved.
    \item GraphWave~\citep{donnat2018learning}: It leverages heat wavelet diffusion patterns to learn a multidimensional structural embedding for each node based on the diffusion of a spectral graph wavelet centered at the node. Then the wavelets as distributions are used to capture structural similarity in graphs.

\end{itemize}
For all baselines, we use the implementation released by the original authors. For our framework \textit{struc2gauss}, we test four variants: \textit{struc2gauss} with expected likelihood and diagonal covariance (\textit{s2g}\_el\_d), expected likelihood and spherical covariance (\textit{s2g}\_el\_s), KL divergence and diagonal covariance (\textit{s2g}\_kl\_d), and KL divergence and spherical covariance (\textit{s2g}\_kl\_s). Note that we only use means of Gaussian distributions as the node embeddings in role clustering and classification tasks. The covariances are left for uncertainty modeling.

For other settings including parameters and evaluation metrics, different settings will be discussed in each task.

\subsection{Case Study: Visualization in 2-D space}
\label{case}
We use the toy example shown in Fig.~\ref{fig:exp} to demonstrate the effectiveness of \textit{struc2gauss} in capturing the global structural information and the failure of other state-of-the-art techniques in this task. The toy network consists of ten nodes and they can be clustered from two different perspectives:
\begin{itemize}
    \item from the perspective of the global role structure, they belong to three groups, i.e., $\{0,1,2,3\}$ (yellow color), $\{4,5,6,7\}$ (blue color) and $\{8,9\}$ (red color) because different groups have different structural functions in this network;
    \item from the perspective of the local community structure, they belong to two groups, i.e., $\{0,1,4,5,6,8\}$ and $\{2,3,6,7,9\}$ because there are denser connections/more edges inside each community that outside the community.
\end{itemize}
Note that from the perspective of role discovery, these three groups of nodes can be explained to play the roles of \textit{periphery}, \textit{star} and \textit{bridge}, respectively.

\begin{figure*}
\centering
    \begin{subfigure}[b]{0.32\textwidth}
        \centering
        \includegraphics[width=\textwidth]{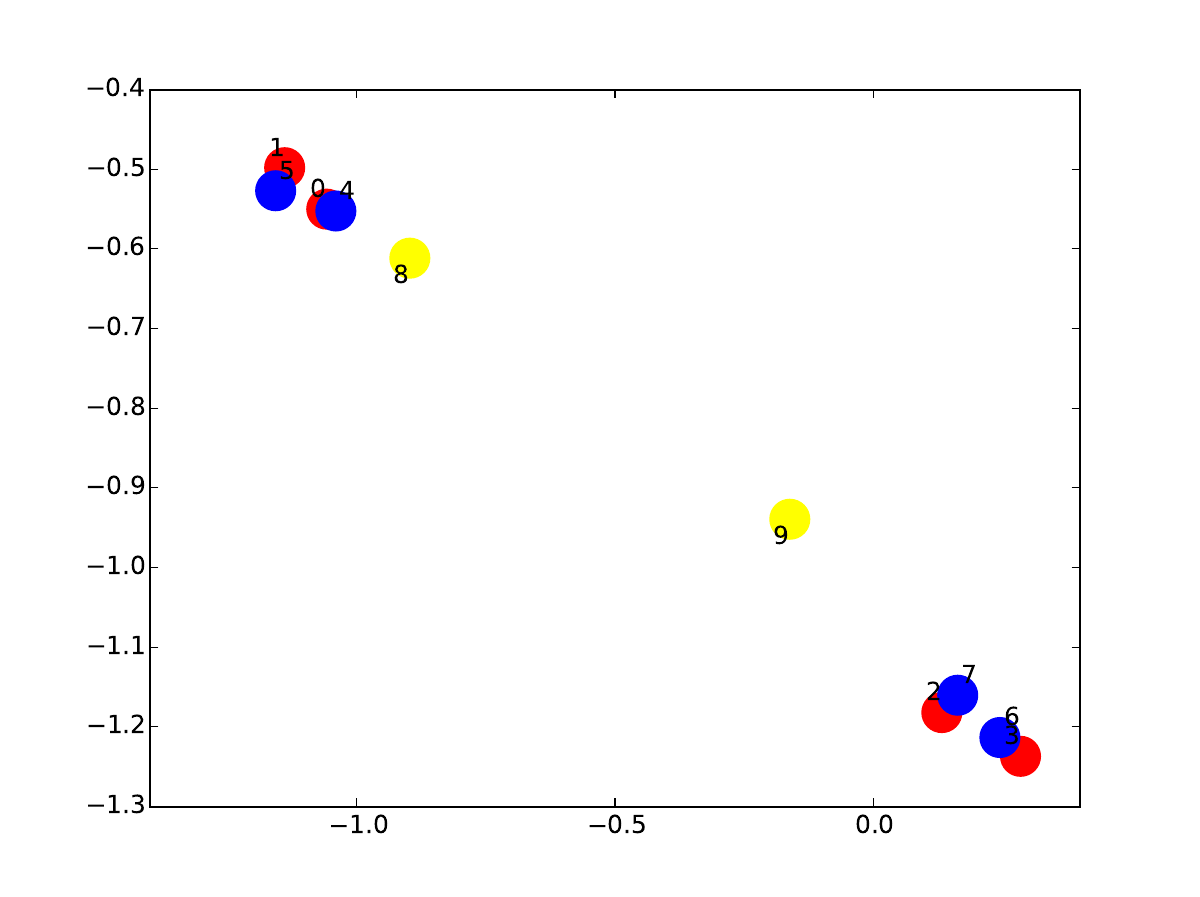}
        \caption[]%
        {{\small DeepWalk.}}    
        \label{fig:a}
    \end{subfigure}
    \hfill
    \begin{subfigure}[b]{0.32\textwidth}  
        \centering 
        \includegraphics[width=\textwidth]{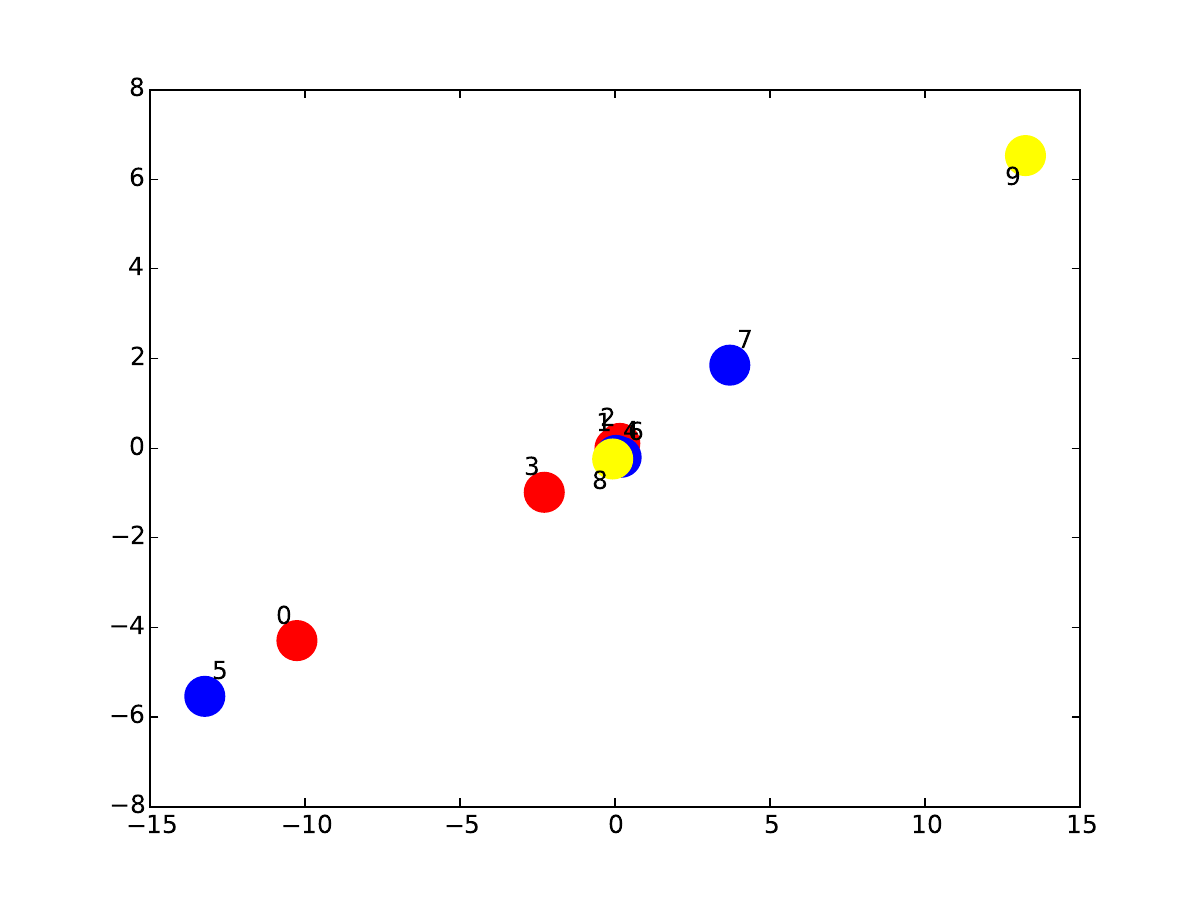}
        \caption[]%
        {{\small LINE}}    
        \label{fig:b}
    \end{subfigure}
    \hfill
    \begin{subfigure}[b]{0.32\textwidth}  
        \centering 
        \includegraphics[width=\textwidth]{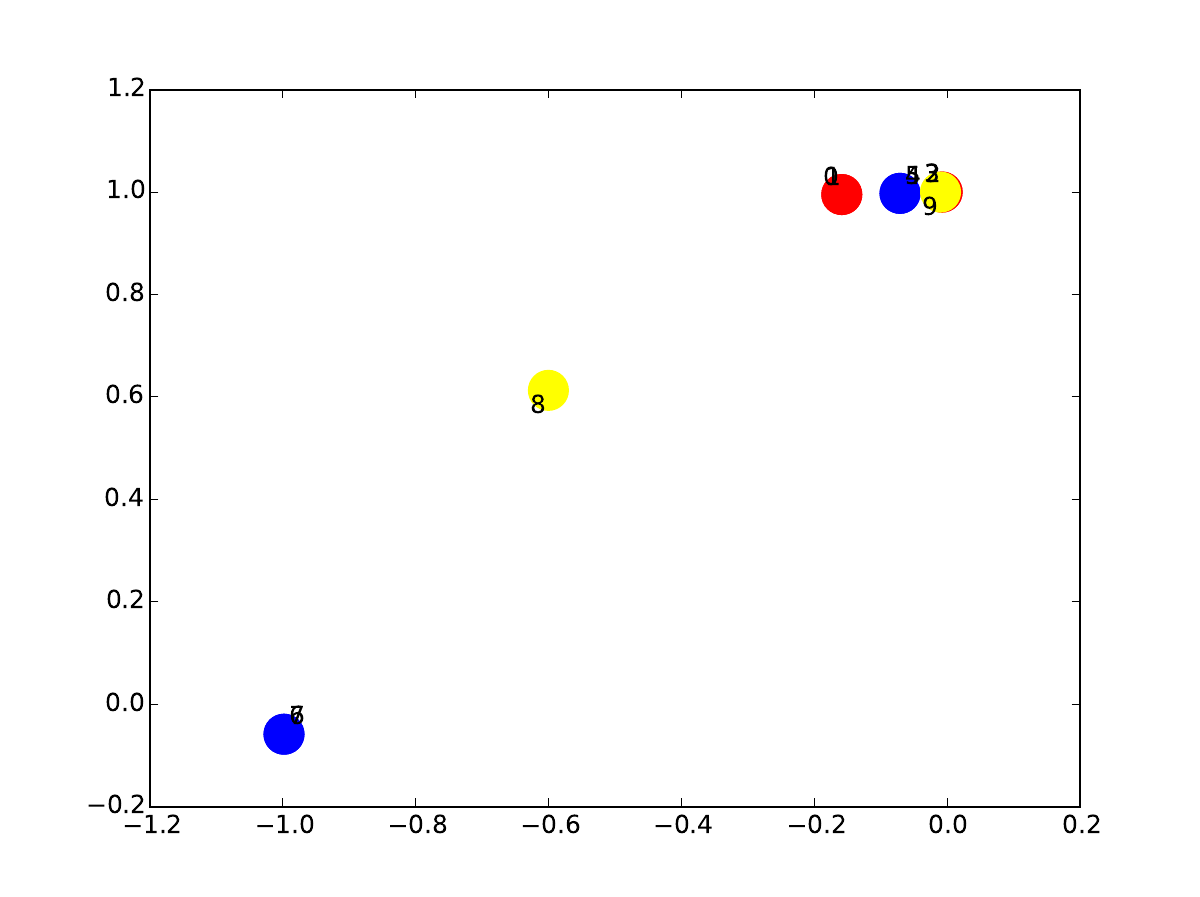}
        \caption[]%
        {{\small \textit{graph2gauss}}}    
        \label{fig:c}
    \end{subfigure}
    \vskip\baselineskip
    \begin{subfigure}[b]{0.32\textwidth}   
        \centering 
        \includegraphics[width=\textwidth]{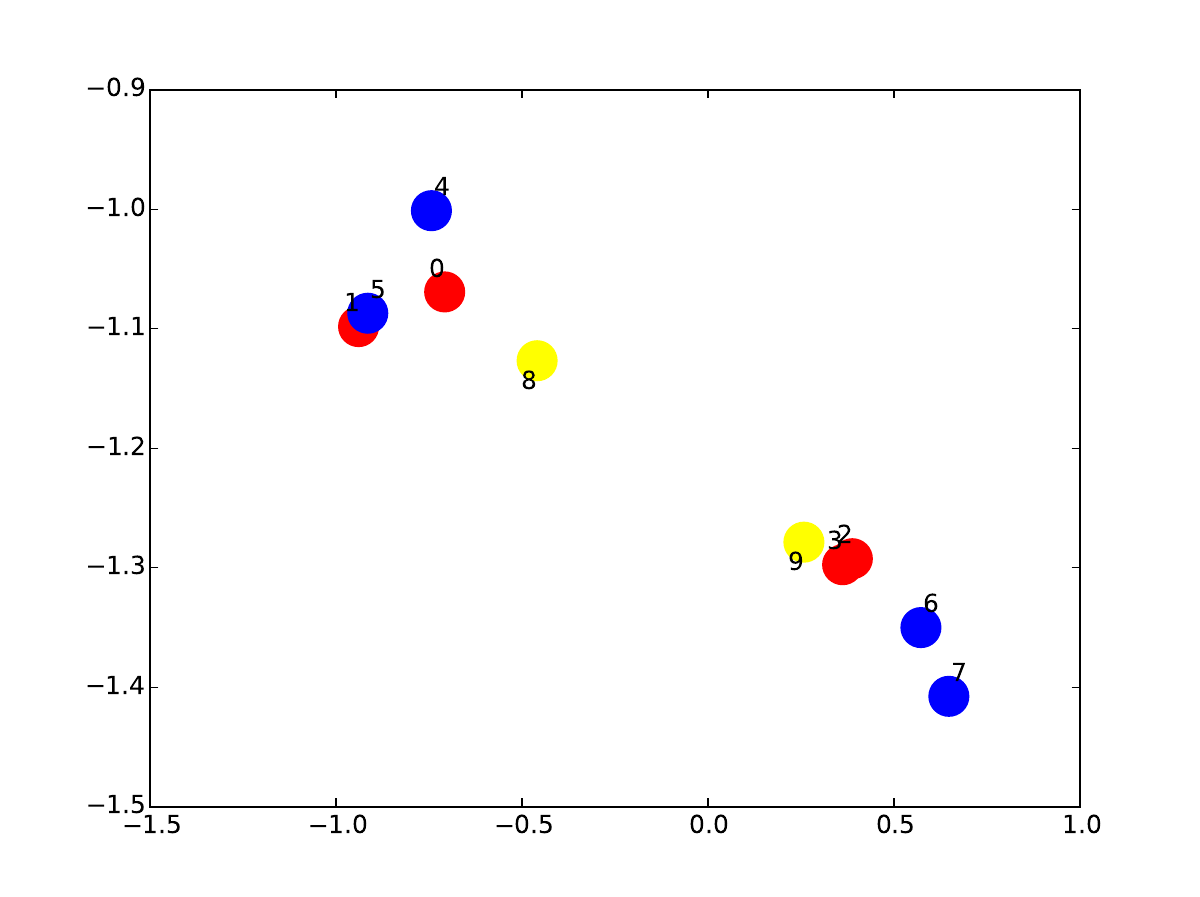}
        \caption[]%
        {{\small \textit{node2vec}}}    
        \label{fig:d}
    \end{subfigure}
    \hfill
    \begin{subfigure}[b]{0.32\textwidth}   
        \centering 
        \includegraphics[width=\textwidth]{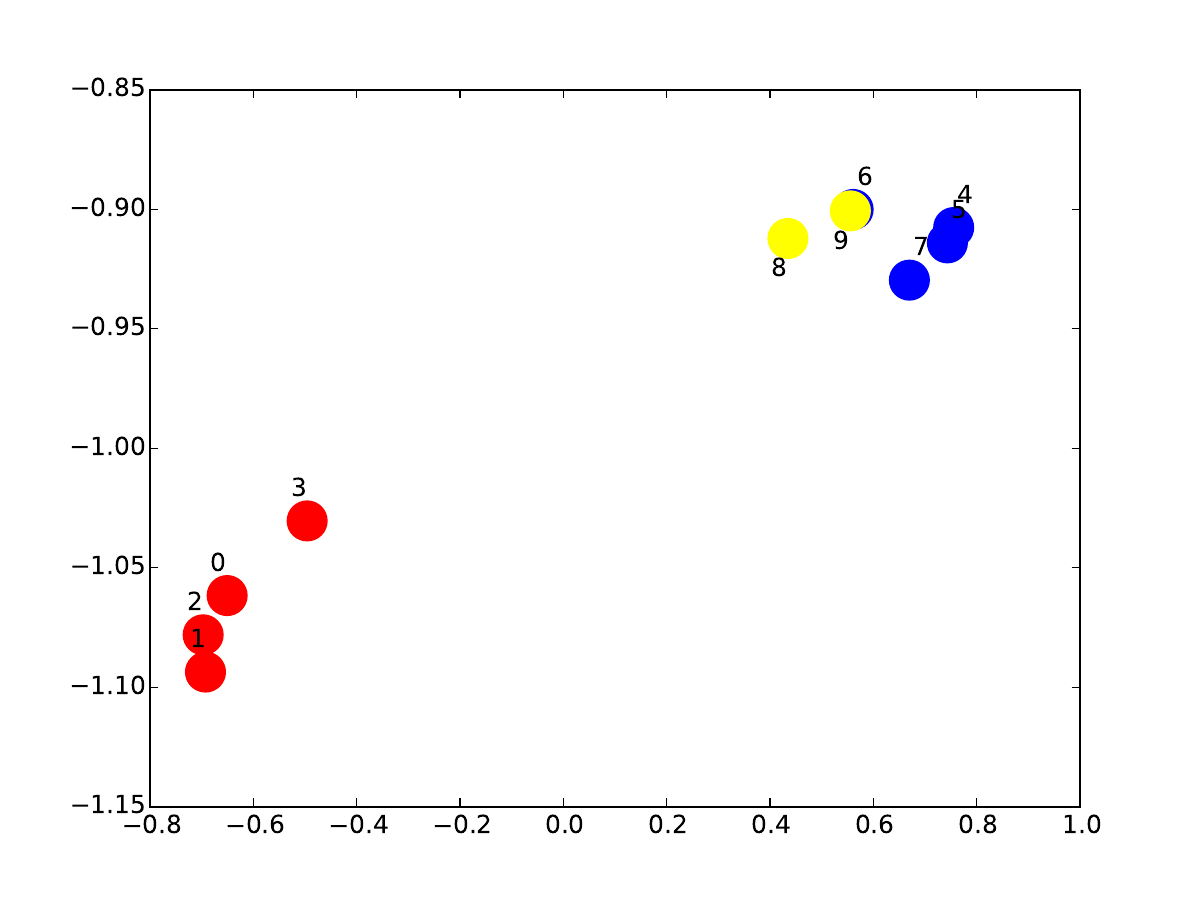}
        \caption[]%
        {{\small \textit{struc2vec}}}    
        \label{fig:e}
    \end{subfigure}
    \hfill
    \begin{subfigure}[b]{0.32\textwidth}   
        \centering 
        \includegraphics[width=\textwidth]{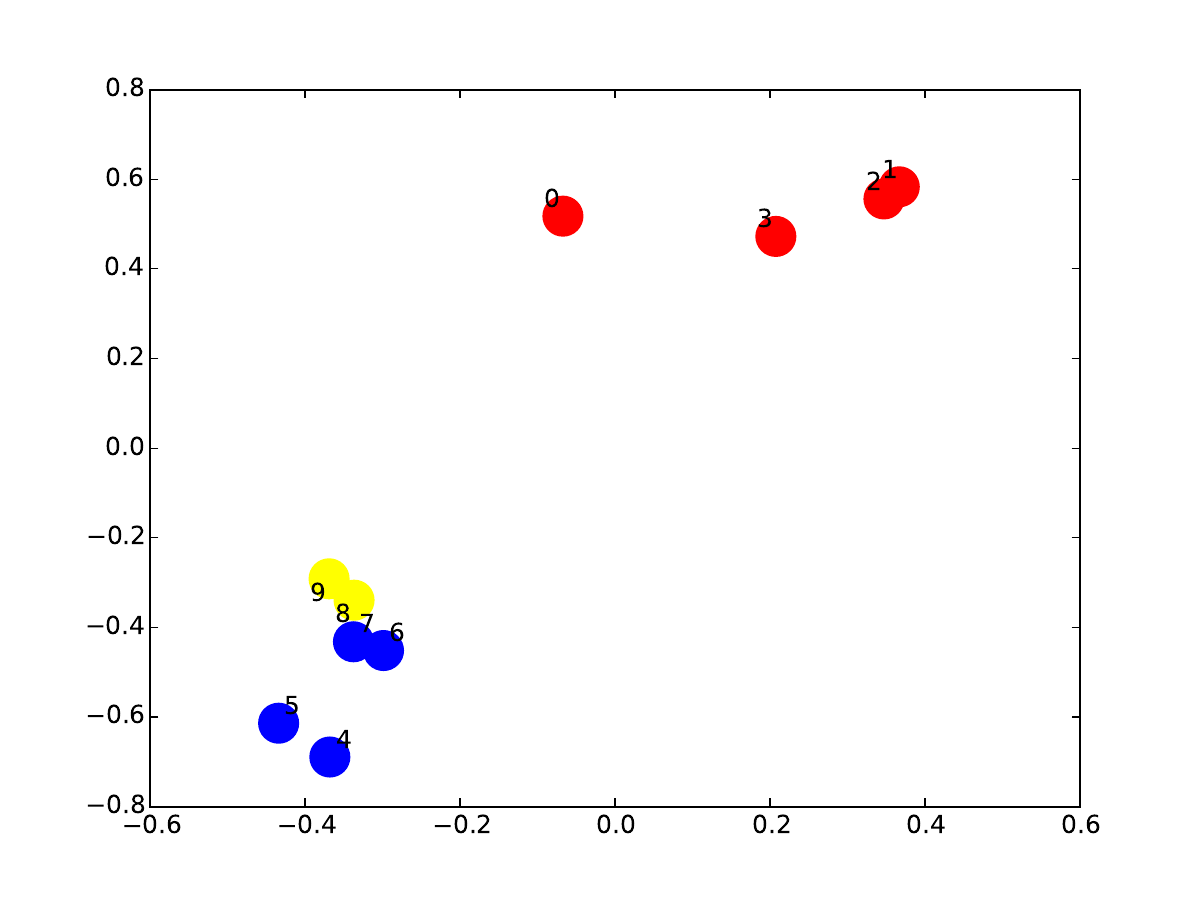}
        \caption[]%
        {{\small \textit{struc2gauss} KL + diag}}    
        \label{fig:f}
    \end{subfigure}
    \vskip\baselineskip
    \begin{subfigure}[b]{0.32\textwidth}   
        \centering 
        \includegraphics[width=\textwidth]{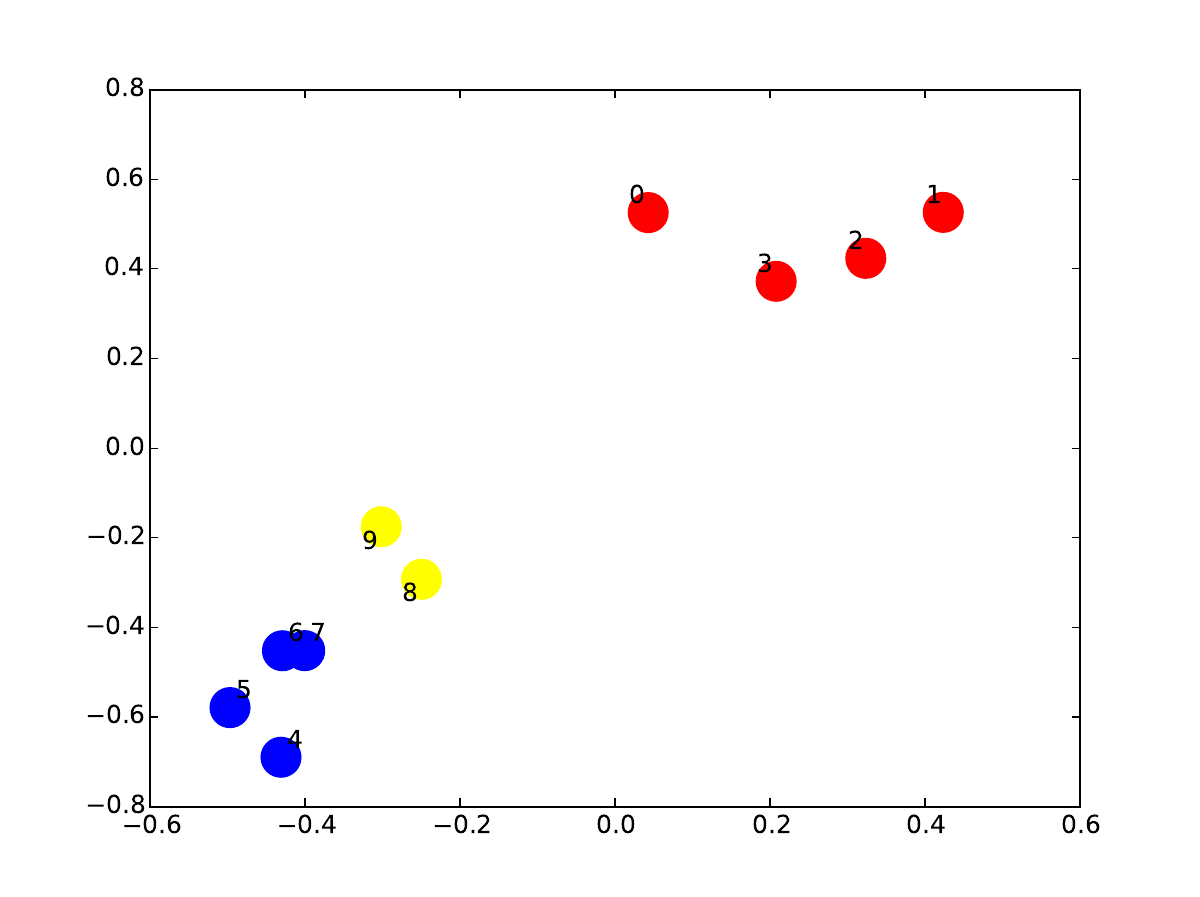}
        \caption[]%
        {{\small \textit{struc2gauss} KL + spher}}    
        \label{fig:g}
    \end{subfigure}
    \hfill
    \begin{subfigure}[b]{0.32\textwidth}   
        \centering 
        \includegraphics[width=\textwidth]{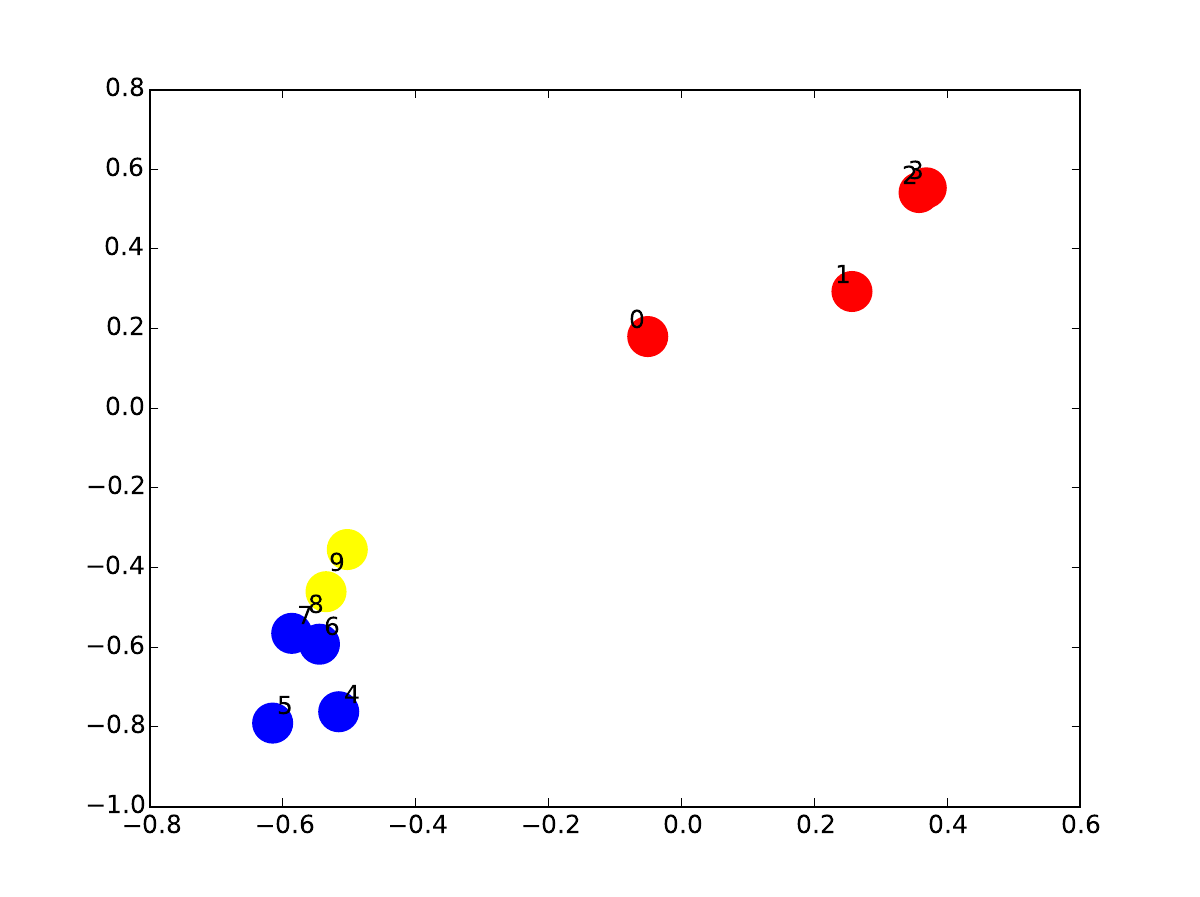}
        \caption[]%
        {{\small \textit{struc2gauss} EL + diag}}    
        \label{fig:h}
    \end{subfigure}
    \hfill
    \begin{subfigure}[b]{0.32\textwidth}   
        \centering 
        \includegraphics[width=\textwidth]{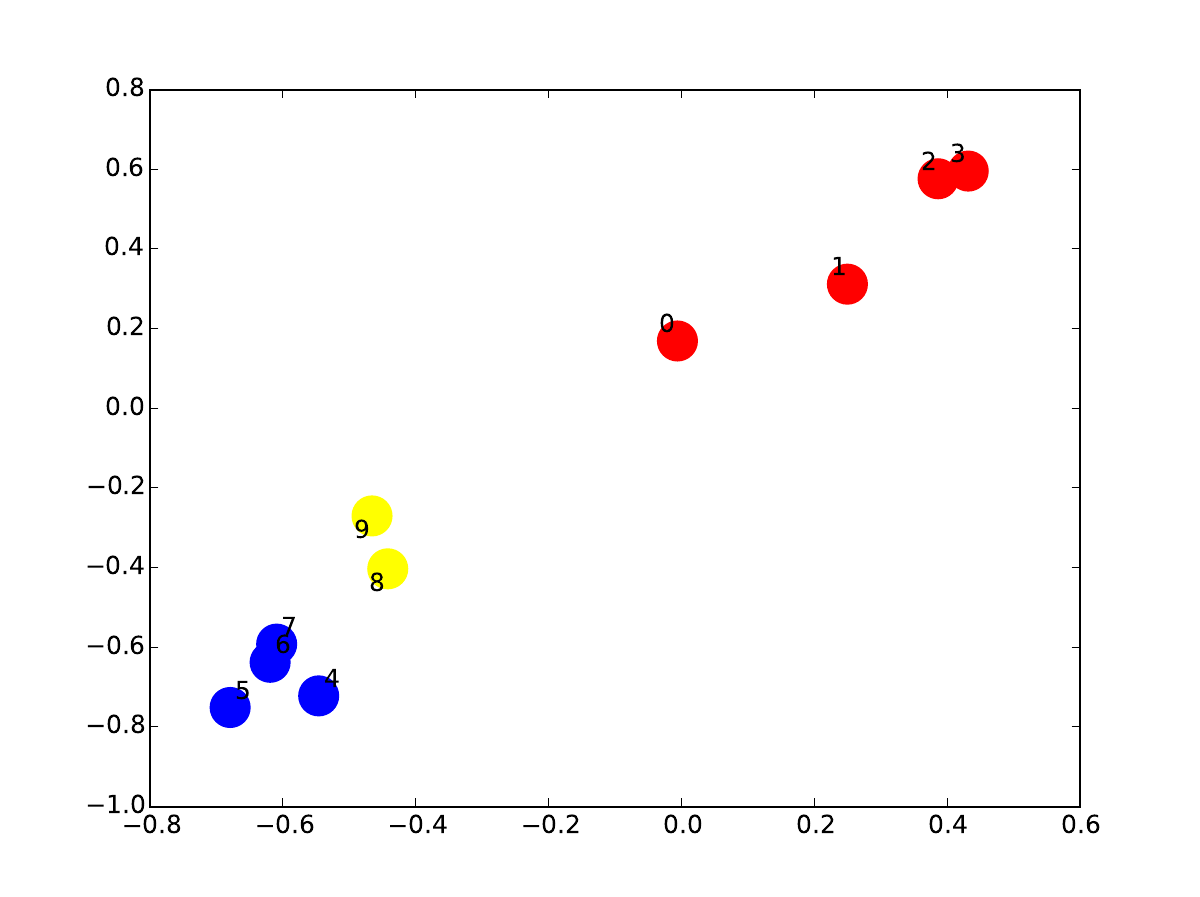}
        \caption[]%
        {{\small \textit{struc2gauss} EL + spher}}    
        \label{fig:i}
    \end{subfigure}
    \caption[]
    {\small  Latent representations in $\mathbb{R}^2$ learned by (a) DeepWalk, (b) LINE, (c) GraRep, (d) \textit{node2vec}, (e) \textit{struc2vec}, (f) \textit{struc2gauss} using KL divergence with diagonal covariance, (g) \textit{struc2gauss} using KL divergence with spherical covariance, (g) \textit{struc2gauss} using KL divergence with diagonal covariance, (h) \textit{struc2gauss} using expected likelihood with diagonal covariance, and (i) \textit{struc2gauss} using expected likelihood with spherical covariance.}
    \label{fig:toy}
\end{figure*}

In this study, we aim to preserve the global structural information in network embedding. Fig.~\ref{fig:toy} shows the learned node representations by different methods. For shared parameters in all methods, we use the same settings by default: representation dimension: 2, number of walks per node: 20, walk length: 80, skipgram window size: 5. For \textit{node2vec}, we set $p = 1$ and $q = 2$. For \textit{graph2gauss} and \textit{struc2gauss}, the number of walks per node is 20 and the number of positive/negative nodes per node is 5. The constraint for means $C$ is 2 and constraints for covariances $c_{min}$ and $c_{max}$ are 0.5 and 2, respectively. From the visualization results, it can be observed that:
\begin{itemize}
    \item Our proposed \textit{struc2gauss} outperforms all other methods. Both diagonal and spherical covariances can separate nodes based on global structural information and \textit{struc2gauss} with spherical covariances performs better than diagonal covariances since it can recognize \textit{star} and \textit{bridge} nodes better.
    \item Methods aim to capture the global structural information performs better than random walk sampling based methods. For example, \textit{struc2vec} can solve this problem to some extent. However, there is overlap between node 6 and 9. It has been stated that \textit{node2vec} can capture the structural equivalence but the visualization shows that it still captures the local structural information similar to DeepWalk.
    \item DeepWalk, LINE and \textit{graph2gauss} fail to capture the global structural information because these methods are based on random walk which only captures the local community structures. DeepWalk is capable to capture the local structural information since nodes are separated into two parts corresponding to the two communities shown in Fig.~\ref{fig:exp}.
\end{itemize}

\subsection{Structural Role Clustering}
\label{cluster}
The most common network mining application based on global structural information is the problem of \textit{role discovery} and role discovery essentially is a clustering task. Thus, we consider this task to illustrate the potential of node representations learned by \textit{struc2gauss}. We use the latent representations learned by different methods (in  \textit{struc2gauss}, we use means of learned Gaussian distribution) as features and K-means as the clustering algorithm to cluster nodes.

\begin{table}
\small
\centering
\caption{NMI for node clustering in air-traffic networks using different NE methods. In \textit{struc2gauss}, el and kl mean expected likelihood and KL divergence, respectively. d and s mean diagonal and spherical covariances, respectively. The highest values are in bold.}
\label{tb:air}
\begin{tabular}{|l|c|c|c|}
\hline
                 & Brazil-air & Europe-air  & USA-air  \\ \hline
DeepWalk~\citep{perozzi2014deepwalk}     & 0.1303         & 0.0458         & 0.0766      \\ \hline
LINE~\citep{tang2015line}   & 0.2215         & 0.1563         & 0.1275     \\ \hline
\textit{node2vec}~\citep{grover2016node2vec}  & 0.2516         & 0.1722         & 0.0945      \\ \hline
EP~\citep{duran2017learning}  & 0.2283         & 0.1405         & 0.1007      \\ \hline
\textit{graph2gauss}~\citep{vilnis2014word}   & 0.1204 & 0.1109 & 0.0896  \\ \hline
\textit{struc2vec}~\citep{ribeiro2017struc2vec}   & 0.3758         & 0.2729         & 0.2486      \\ \hline
DRNE~\citep{tu2018deep} & 0.5244  & 0.2766  &  0.2918 \\ \hline
GraphWave~\citep{donnat2018learning} & 0.5040  & 0.3230  &  0.2452 \\ \hline
\textit{s2g}\_el\_d      & 0.5615         & 0.3234         & 0.3188      \\ \hline
\textit{s2g}\_el\_s      & 0.5396         & 0.2974         & 0.2967      \\ \hline
\textit{s2g}\_kl\_d      & 0.5527         & 0.3145         & 0.3212      \\ \hline
\textit{s2g}\_kl\_s      & \textbf{0.5675}         & \textbf{0.3280}         & \textbf{0.3217}      \\ \hline
\end{tabular}
\end{table}

\begin{figure}
\centering
\includegraphics[width=4.2in]{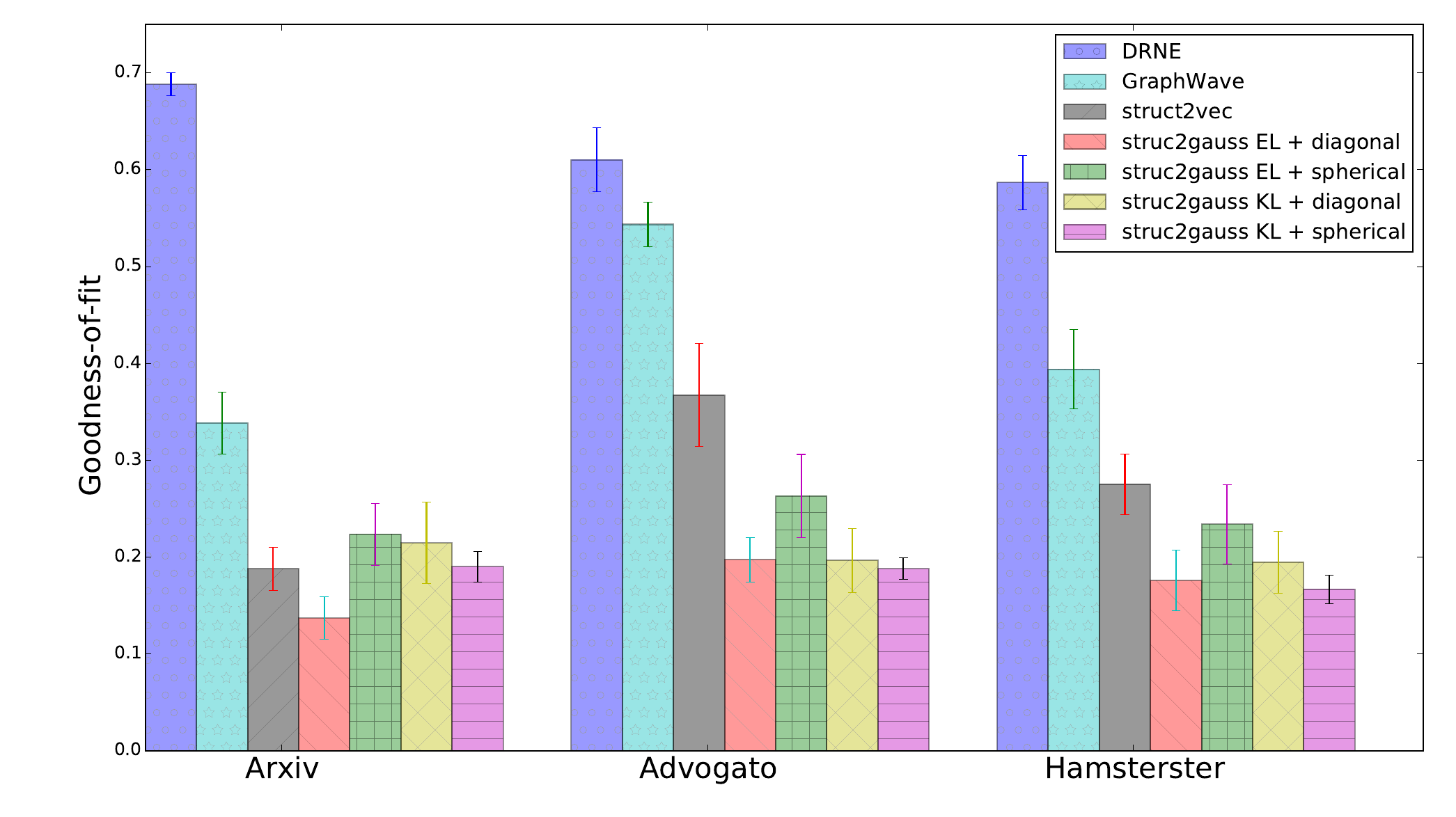}
\caption{Goodness-of-fit of global structure preserving embedding baselines and \textit{struc2gauss} with different strategies on three real-world networks. Lower value means better performance.}
\label{fig:gof}
\end{figure}

\textbf{Parameters}. For these baselines, we use the same settings in the literature: representation dimension: 128, number of walks per node: 20, walk length: 80, skipgram window size: 10. For \textit{node2vec}, we set $p = 1$ and $q = 2$. For \textit{graph2gauss} and \textit{struc2gauss}, we set the constraint for means $C$ to be 2 and constraints for covariances $c_{min}$ and $c_{max}$ to be 0.5 and 2, respectively. The number of walks per node is 10, the number of positive/negative nodes per node is 120 and the representation dimension is also 128.

\textbf{Evaluation Metrics}. To quantitatively evaluate clustering performance in labeled networks, we use \textit{Normalized Mutual Information (NMI)} as the evaluation metric. NMI is obtained by dividing the mutual information by the arithmetic average of the entropy of obtained cluster and ground-truth cluster. It evaluates the clustering quality based on information theory, and is defined by normalization on the mutual information between the cluster assignments and the pre-existing input labeling of the classes:
\begin{equation}
\label{equation:NMI}
    NMI(\mathcal{C,D})=\frac{2*\mathcal{I(C,D)}}{\mathcal{H(C)+H(D)}},
\end{equation}
where obtained cluster $\mathcal{C}$ and ground-truth cluster $\mathcal{D}$. The mutual information $\mathcal{I}(\mathcal{C},\mathcal{D})$ is defined as $\mathcal{\mathcal{I}}(\mathcal{C},\mathcal{D})=\mathcal{H}(\mathcal{C})-\mathcal{H(C|D)}$ and $\mathcal{H}(\cdot)$ is the entropy.

For unlabeled networks, we use normalized \textit{goodness-of-fit} as the evaluation metric. \textit{goodness-of-fit} can measure how well the representation of roles and the relations among these roles fit a given network~\citep{wasserman1994social}. In \textit{goodness-of-fit}, it is assumed that the output of a role discovery method is an optimal model, and nodes belonging to the same role are predicted to be perfectly structurally equivalent. In real-world social networks, nodes belonging to the same role are only approximately structurally equivalent. The essence of \textit{goodness-of-fit indices} is to measure how just how approximate are the approximate structural equivalences. If the optimal model holds, then all nodes belonging to the same role are exactly structurally equivalent.

In detail, given a social network with $n$ vertices $V=\{v_1,v_2,...,v_n\}$ and $m$ roles, we have the adjacency matrix $A=\{A_{ij}\in \{0,1\}|1\leq i,j\leq n\}$ and the role set $R=\{R_1,R_2,...,R_m\}$, where $v_i\in R_j$ indicates node $v_i$ belongs to the $j$th role, as obtained using DyNMF. Note that $R$ partitions $V$, in the sense that each $v\in V$ belongs to exactly one $R_i\in R$. Then the density matrix $\Delta$ is defined as:
\begin{equation}
\Delta_{ij}=
\begin{cases}
    \sum_{v_k\in R_i,v_l\in R_j}A_{kl}/(|R_i|\cdot|R_j|),   &  \text{if~} i\neq j \\
    \sum_{v_k\in R_i,v_l\in R_j}A_{kl}/(|R_i|\cdot(|R_j|-1)),   &  \text{if~} i=j.
\end{cases}
\end{equation}
We also define block matrix $B$ based on the discovered roles. In fact, there are several criteria which can be used to build the block matrix including perfect fit, zeroblock, oneblock and $\alpha$ density criterion~\citep{wasserman1994social}. Since real social network data rarely contain perfectly structural equivalent nodes~\citep{faust1992blockmodels}, perfect fit, zeroblock and oneblock criteria would not work well in real-world data and we use $\alpha$ density criterion to construct the block matrix $B$:
\begin{equation}
B_{ij}=
\begin{cases}
    0,   &  \text{if~} \Delta_{ij}<\alpha \\
    1,   &  \text{if~} \Delta_{ij}\geq\alpha
\end{cases}
\end{equation}
where $\alpha$ is the threshold to determine the values in blocks. $\alpha$ density criterion is based on the density of edges between nodes belong to the same role and defined as
\begin{equation}
\alpha=\sum_{1\leq i,j\leq n}A_{ij}/(n(n-1)).
\end{equation}
Based on the definitions of density matrix $\Delta$ and block matrix $B$, the \textit{goodness-of-fit index} $e$ is defined as
\begin{align}
    e=\sum_{1\leq i,j\leq m}|B_{ij}-\Delta_{ij}|.
\end{align}
To make the evaluation metric value in the range of $[0,1]$, we normalize \textit{goodness-of-fit} by dividing $r^2$ where $r$ is number of groups/roles. For more details about \textit{goodness-of-fit indices}, please refer to~\citep{wasserman1994social}.

\textbf{Results}. The NMI values for node clustering on networks with labels are shown in Table~\ref{tb:air} and the normalized \textit{goodness-of-fit} values for networks without labels are shown in Fig.~\ref{fig:gof}. Note that random walk and neighbor based embedding methods, including DeepWalk, LINE, \textit{node2vec}, EP and \textit{graph2gauss}, aim at capturing local structural information and so are incapable of preserving structural roles. Hence,  for simplicity, we will not compare them to these role preserving methods on networks without clustering labels.

From these results, some conclusions can be drawn:
\begin{itemize}
    \item For both types of networks with and without clustering labels, \textit{struc2gauss} outperforms all other methods in different evaluation metrics. It indicates the effectiveness of \textit{struc2gauss} in capturing the global structural information.
    \item Comparing \textit{struc2gauss} with diagonal and spherical covariances, it can be observed that spherical covariance can achieve better performance in node clustering. This finding is similar to the results of word embedding in~\citep{vilnis2014word}. A possible explanation could be: spherical covariance requires the diagonal elements to be the same which limits the representation power of covariance matrices but on the contrast enhance the representation power of the learned means. Since we only use means to represent nodes, the method with spherical covariance matrix could learn more relaxed means which leads to better performance. 
    \item For baselines, \textit{struc2vec}, GraphWave and DRNE can capture the structural role information to some extent since their performance is better than these random walk based methods, i.e., DeepWalk and \textit{node2vec}, and neighbor-based method, i.e., EP and \textit{graph2gauss}, while all of them fail in capturing the global structural information for node clustering.
\end{itemize}

\subsection{Structural Role Classification}
\label{classification}
Node classification is another widely used task for embedding evaluation. Different from previous studies which focused on community structures, our approach aims to preserve the global role structures. Thus, we evaluate the effectiveness of \textit{struc2gauss} in role classification task. Same to the node clustering task in Section~\ref{cluster}, we use the latent representations learned by different methods as features. Each dataset is separated into training set and  test set (we will explore the classification performance with different percentages of training set). To focus on the learned representation, we use logistic regression as the classifier.
 
Structural role classification as a supervised task, the ground-truth labels are required. Thus we only use two air-traffic networks for evaluation. We compare our approach to the same state-of-the-art NE algorithms as baselines used in Section~\ref{cluster}, i.e., DeepWalk, LINE, \textit{node2vec}, EP, \textit{graph2gauss}, \textit{struc2vec}, GraphWave and DRNE. Same to~\citep{tu2018deep}, we also compare to four centrality measures, i.e., closeness centrality, betweenness centrality, eigenvector centrality and k-core. Since the combination of these four measures perform best~\citep{tu2018deep}, we only compare the classification performance of the combination as features in this task. The parameters of baselines and \textit{struc2gauss}, we use the same settings in Section~\ref{cluster}.

\begin{figure}
\centering
\includegraphics[width=3.8in]{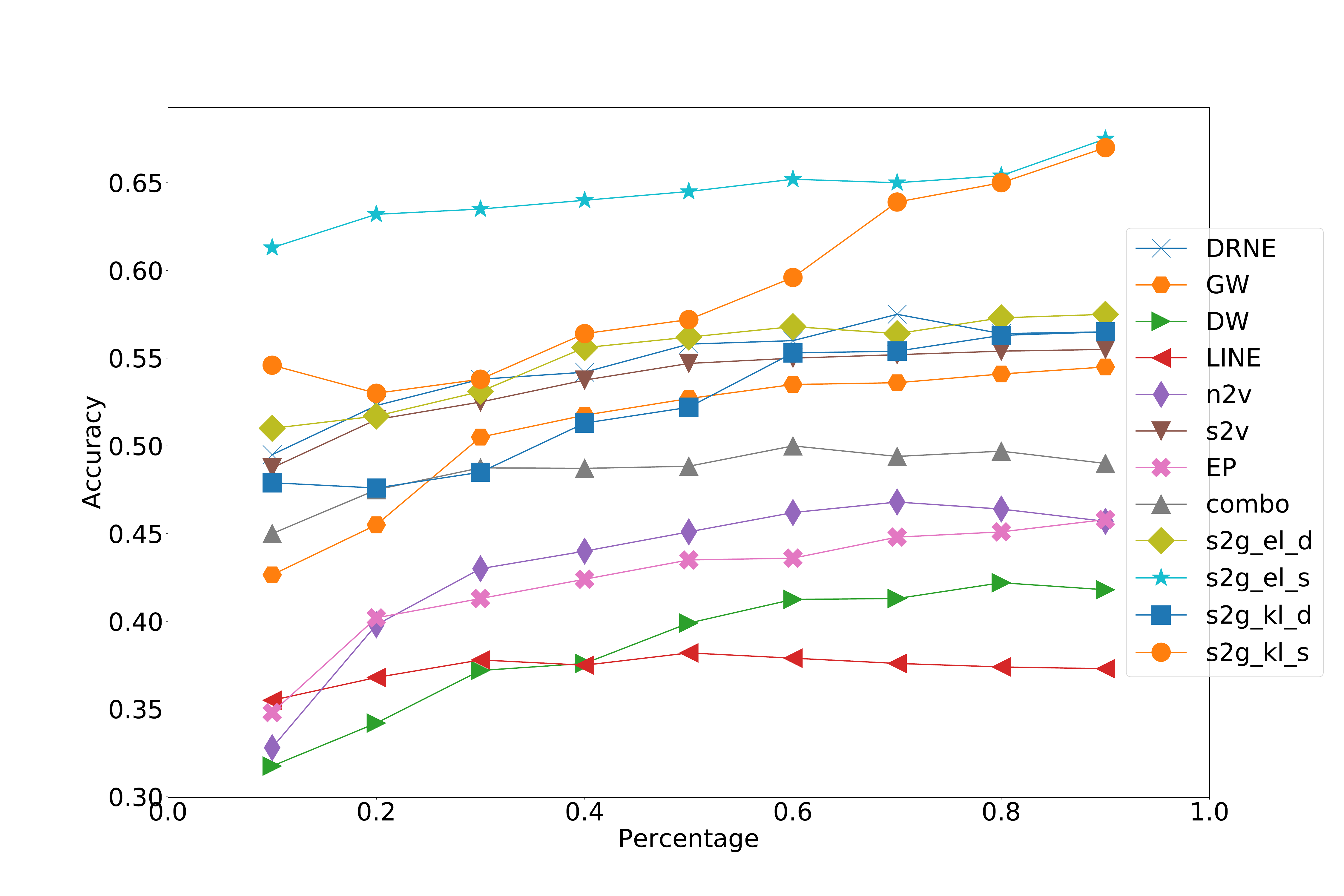}
\caption{Average accuracy for structural role classification in Europe-air network.}
\label{fig:acc1}
\end{figure}

\begin{figure}
\centering
\includegraphics[width=3.8in]{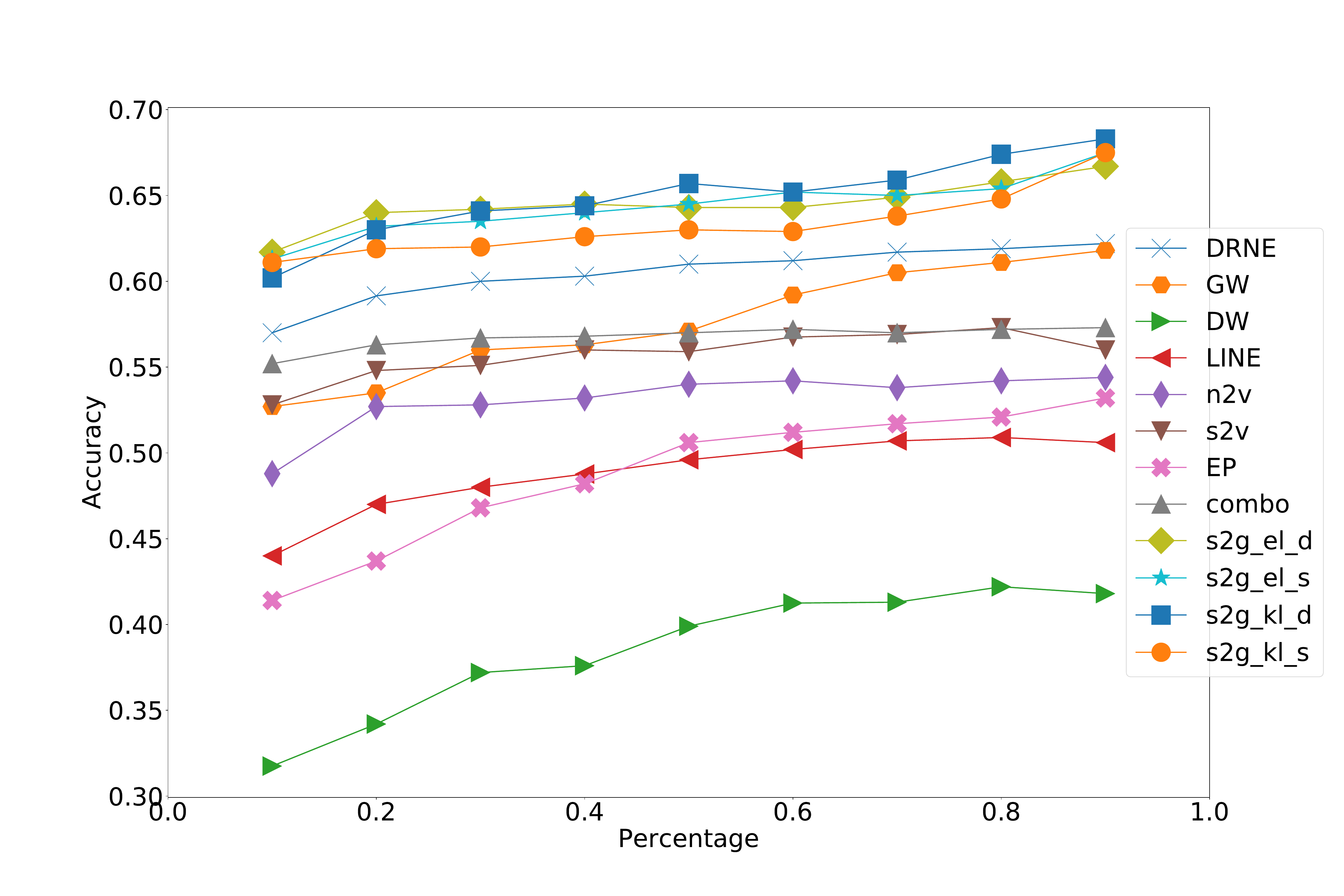}
\caption{Average accuracy for structural role classification in USA-air network}
\label{fig:acc2}
\end{figure}

The average accuracies for structural role classification in Europe-air and USA-air are shown in Fig.~\ref{fig:acc1} and~\ref{fig:acc2}. From the results, we can observe that:
\begin{itemize}
    \item \textit{struc2gauss} outperforms almost all other methods in both networks except DRNE in Europe-air network.
    In Europe-air network, \textit{struc2gauss} with expected likelihood and spherical covariances, i.e., s2g\_el\_s, performs best. \textit{struc2gauss} with KL divergence and spherical covariances, i.e., s2g\_kl\_s, achieves the second best performance especially when the training ratio is larger than 0.7. \textit{struc2gauss} with diagonal covariances, i.e., s2g\_el\_d and s2g\_kl\_d, are on par with GraphWave, DRNE and \textit{struc2vec} and outperform other methods. In the USA-air network, \textit{struc2gauss} with different settings outperforms all baselines. This indicates the effectiveness of \textit{struc2gauss} in modeling the structural role information. Although not the same combination of energy function and covariance form performs best in two networks, different variants of \textit{struc2gauss} are always the best.
    \item Among the baselines, only \textit{struc2vec}, GraphWave and DRNE can capture the structural role information so that they achieve better classification accuracy than other baselines. DRNE performs the best among these baselines since it captures regular equivalence. GraphWave and \textit{struc2vec} are the second best baselines because they also aim to capture structural roles.
    \item Random walk and neighbor based NE methods only capture local community structures so they perform worse than \textit{struc2gauss}, GraphWave, DRNE and our proposed \textit{struc2gauss}. Node that methods such as DeepWalk, LINE and \textit{node2vec}, although considering the first-, second- and/or higher-order proximity, still are not capable of modeling structural role information.
\end{itemize}


\subsection{Uncertainty Modeling}
Mapping a node in a network into a distribution rather than a point vector allows us to model the uncertainty of the learned representation which is another advantage of \textit{struc2gauss}. Different factors can lead to uncertainties of data. It is intuitive that the more noisy edges a node has, the less discriminative information it contains, thus making its embedding more uncertain. Similarly, incompleteness of information in the network can also bring uncertainties to the representation learning. Therefore, in this section, we study two factors: \textit{noisy information} and \textit{incomplete information}. 

To verify these hypotheses, we conduct the following experiment using Brazil-air and Europe-air networks. For \textit{noisy information}, we randomly insert certain number of edges to the network and then learn the latent representations and covariances. The average variance is used to measure the uncertainties. For Brazil-air network, we range the number of noisy edges from 50 to 300 and for Europe-air it ranges from 500 to 3000. For \textit{incomplete information}, we randomly delete certain number of edges to the network to make it incomplete and then learn the latent representations and covariances. Similarly, for Brazil-air network, we range the number of removed edges from 50 to 300 and for Europe-air it ranges from 500 to 3000. The other parameter settings are same to Section~\ref{cluster}. 

\begin{figure}
\centering
    \begin{subfigure}[b]{0.85\textwidth}
        \centering
        \includegraphics[width=2.8in]{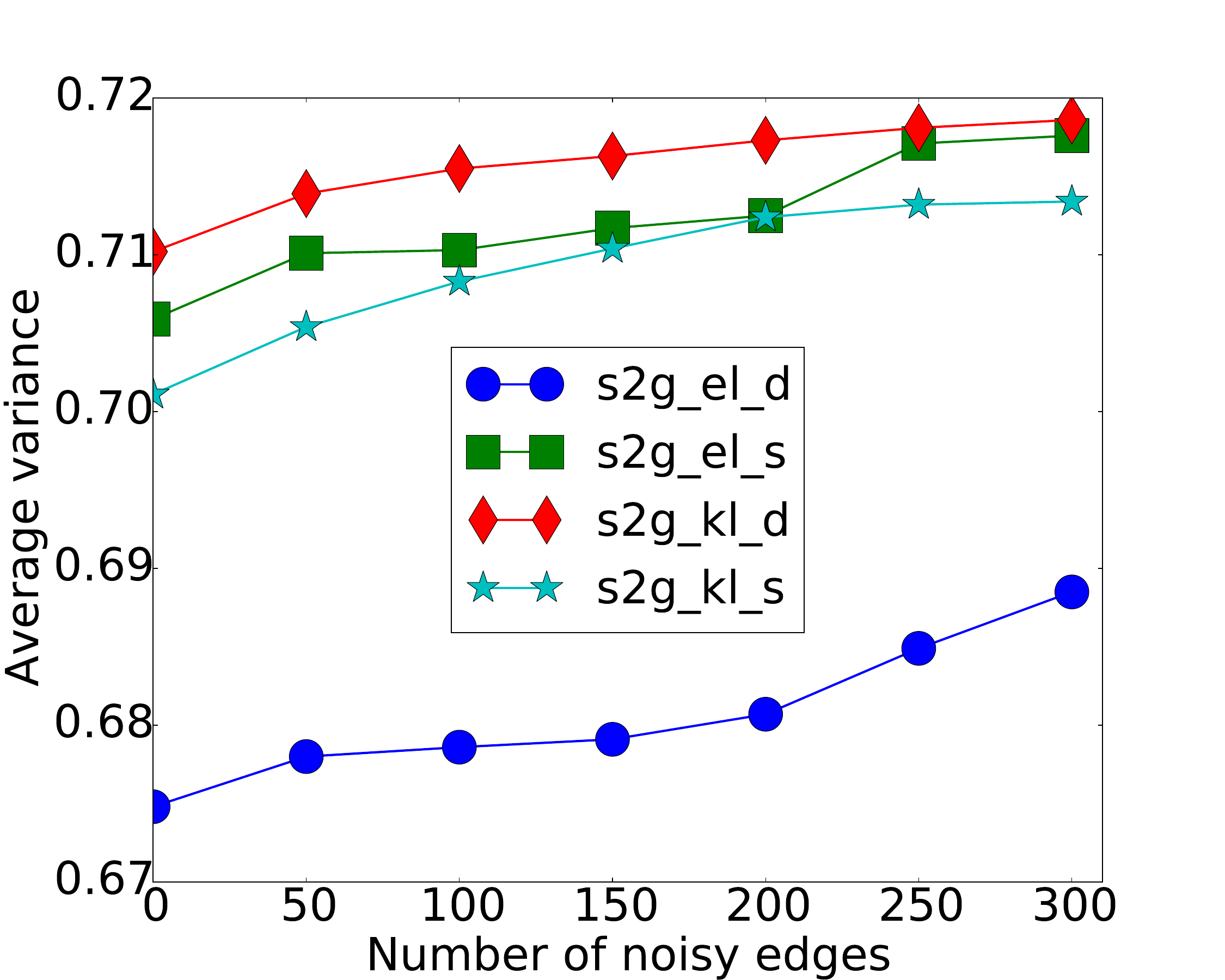}
        \caption[]%
        {{\small Average variance with different numbers of noisy edges on Brazil-air.}}
        \label{fig:unc1}
    \end{subfigure}
    \vskip\baselineskip
    \begin{subfigure}[b]{0.85\textwidth}
        \centering
        \includegraphics[width=2.8in]{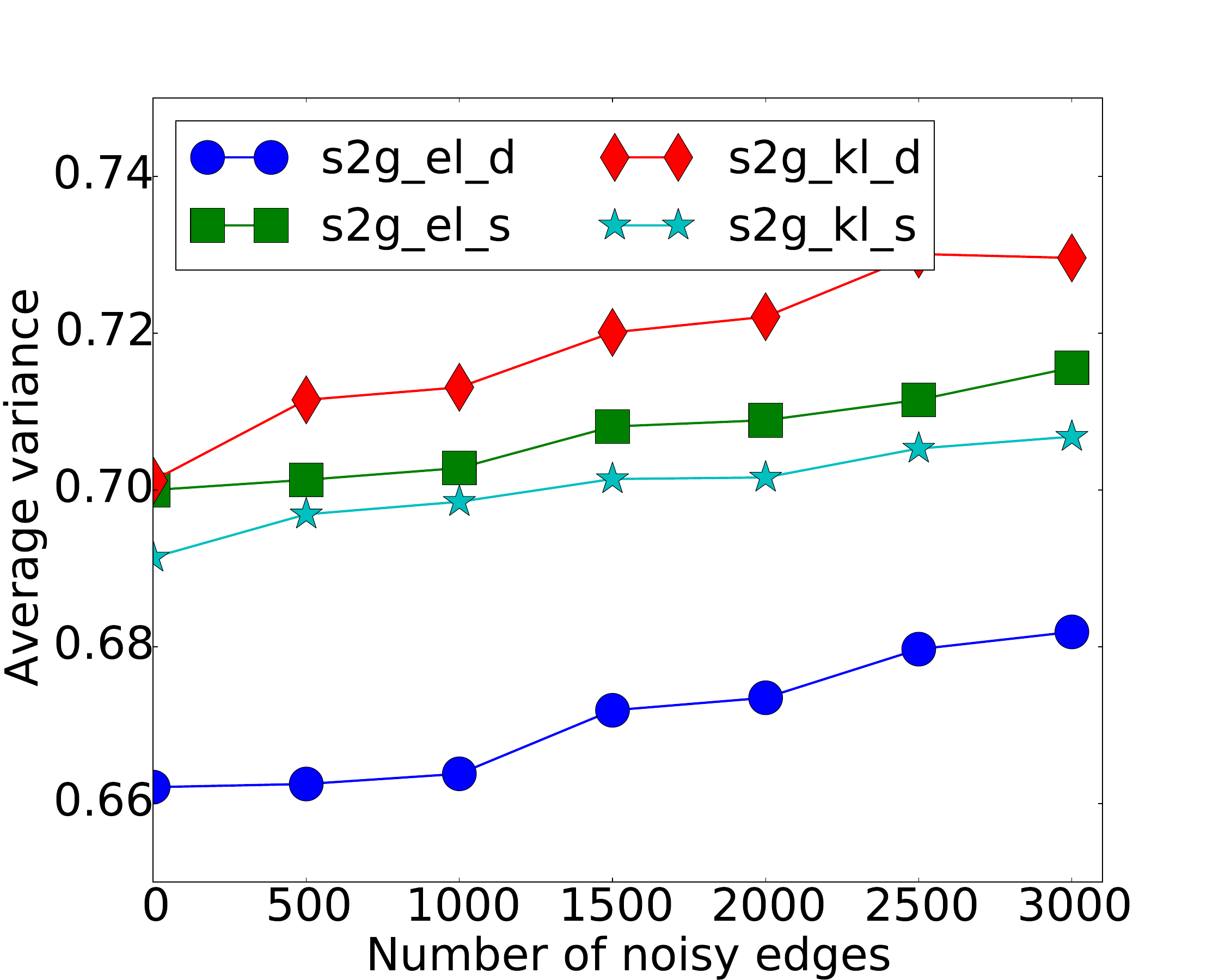}
        \caption[]%
        {{\small Average variance with different numbers of noisy edges on Europe-air.}}
        \label{fig:unc2}
    \end{subfigure}
    \caption{Uncertainties of embeddings with different levels of noise.}
    \label{fig:unc}
\end{figure}

\begin{figure}
\centering
    \begin{subfigure}[b]{0.85\textwidth}
        \centering
        \includegraphics[width=2.8in]{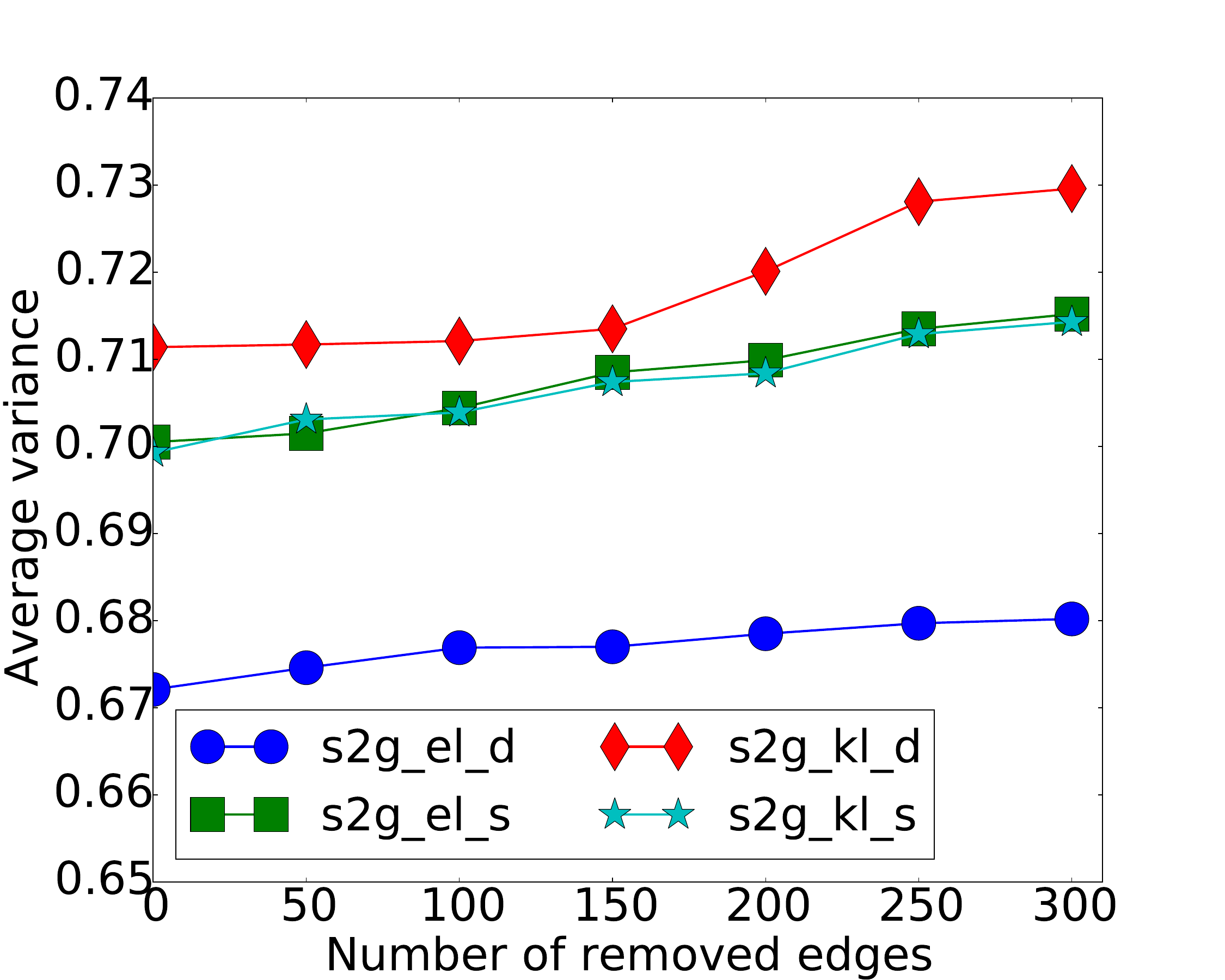}
        \caption[]%
        {{\small Average variance with different numbers of removed edges on Brazil-air.}}
        \label{fig:inc1}
    \end{subfigure}
    \vskip\baselineskip
    \begin{subfigure}[b]{0.85\textwidth}
        \centering
        \includegraphics[width=2.8in]{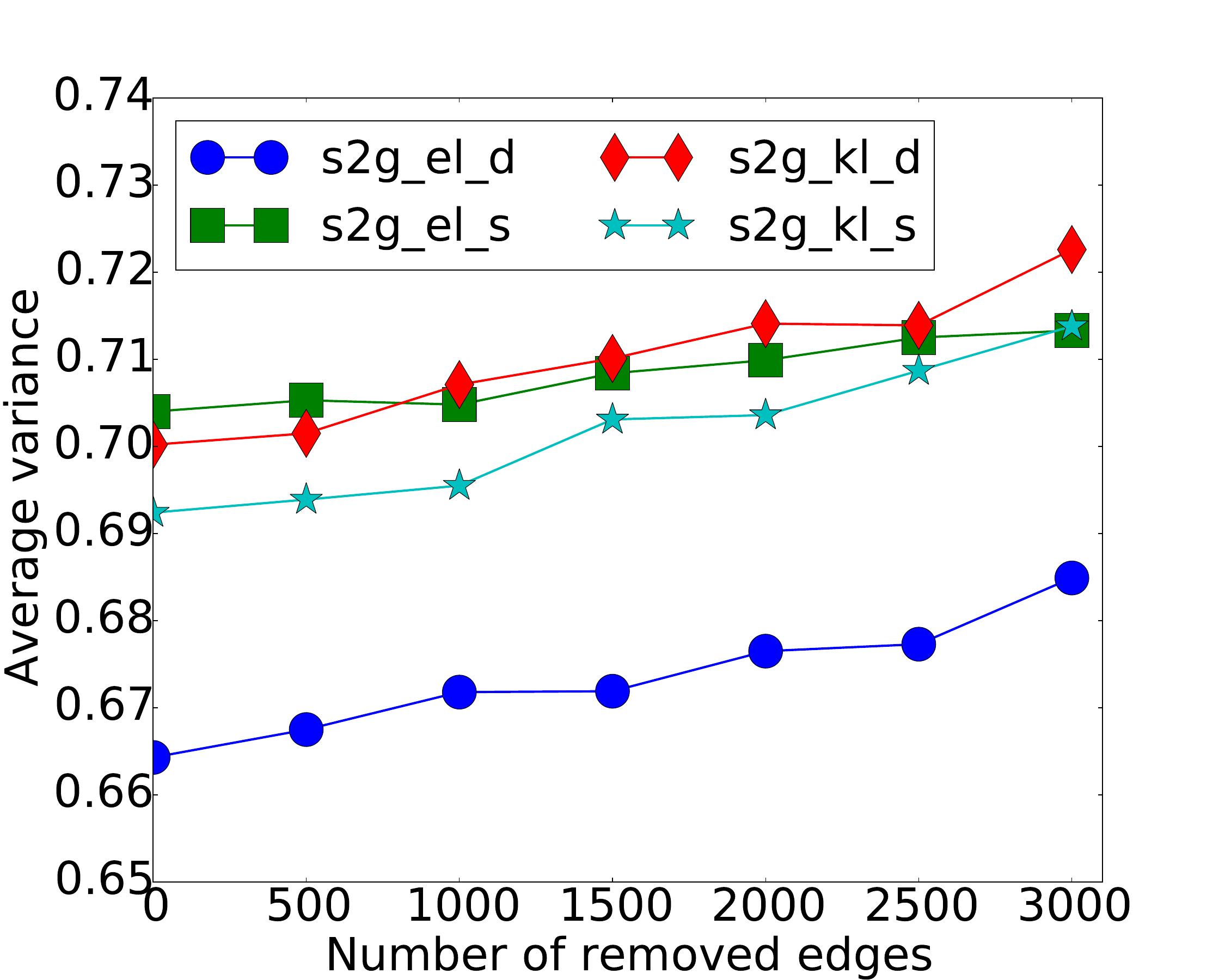}
        \caption[]%
        {{\small Average variance with different numbers of removed edges on Europe-air.}}
        \label{fig:inc2}
    \end{subfigure}
    \caption{Uncertainties of embeddings with different levels of incompleteness.}
    \label{fig:inc}
\end{figure}

The results are shown in Fig.~\ref{fig:unc} and Fig.~\ref{fig:inc}. It can be observed that (1) with more noisy edges being added to the networks and (2) with more removed edges from the networks, average variance values become larger. \textit{struc2gauss} with different energy functions and covariance forms have the same trend. This demonstrates that our proposed \textit{struc2gauss} is able to model the uncertainties of learned node representations. It is interesting that \textit{struc2gauss} with expected likelihood and diagonal covariance (s2g\_el\_d) always has the lowest average variance while \textit{struc2gauss} with KL divergence and diagonal (s2g\_kl\_d) always has the largest value. This may result from the learning mechanism of different energy functions when measuring the distance between two distributions. To clarify the results, we also list the NMI for the clustering task in Table~\ref{tb:noise1} and~\ref{tb:noise2}. Compared to the original Gaussian embedding method, we again show the effectiveness of our method in preserving structural role and modeling uncertainties.

\subsection{Influence of Similarity Measures}
\label{sim}
As we mentioned not all structural similarity measures can capture the global structural role information, to validate the rationale to select RoleSim as the similarity measure for structural role information, we investigate the influence of different similarity measures on learning node representations. In specific, we select two other widely used structural similarity measures, i.e., SimRank~\citep{jeh2002simrank} and MatchSim~\citep{lin2009matchsim}, and we incorporate these measures by replacing RoleSim in our framework. The datasets and evaluation metrics used in this experiment are the same to Section~\ref{cluster}. For simplicity, we only show the results of \textit{struc2gauss} using KL divergence with spherical covariance in this experiment because different variants perform similarly in previous experiments.
\begin{table}
\small
\centering
\caption{NMI for node clustering in air-traffic networks of Brazil, Europe and USA using \textit{struc2gauss} with different similarity measures.}
\label{tb:sim}
\begin{tabular}{|l|c|c|c|}
\hline
         & Brazil-air     & Europe-air     & USA-air     \\ \hline
SimRank  & 0.1695 $\pm~0.013$        & 0.0524 $\pm~0.009$        & 0.0887 $\pm~0.025$     \\ \hline
MatchSim & 0.3534 $\pm~0.024$        & 0.2389 $\pm~0.020$        & 0.0913 $\pm~0.013$     \\ \hline
RoleSim  & \textbf{0.5675 $\pm$ 0.032}        & \textbf{0.3280 $\pm$ 0.019}        & \textbf{0.3217 $\pm$ 0.023}     \\ \hline
\end{tabular}
\end{table}
\begin{figure}
\centering
\includegraphics[width=2.8in]{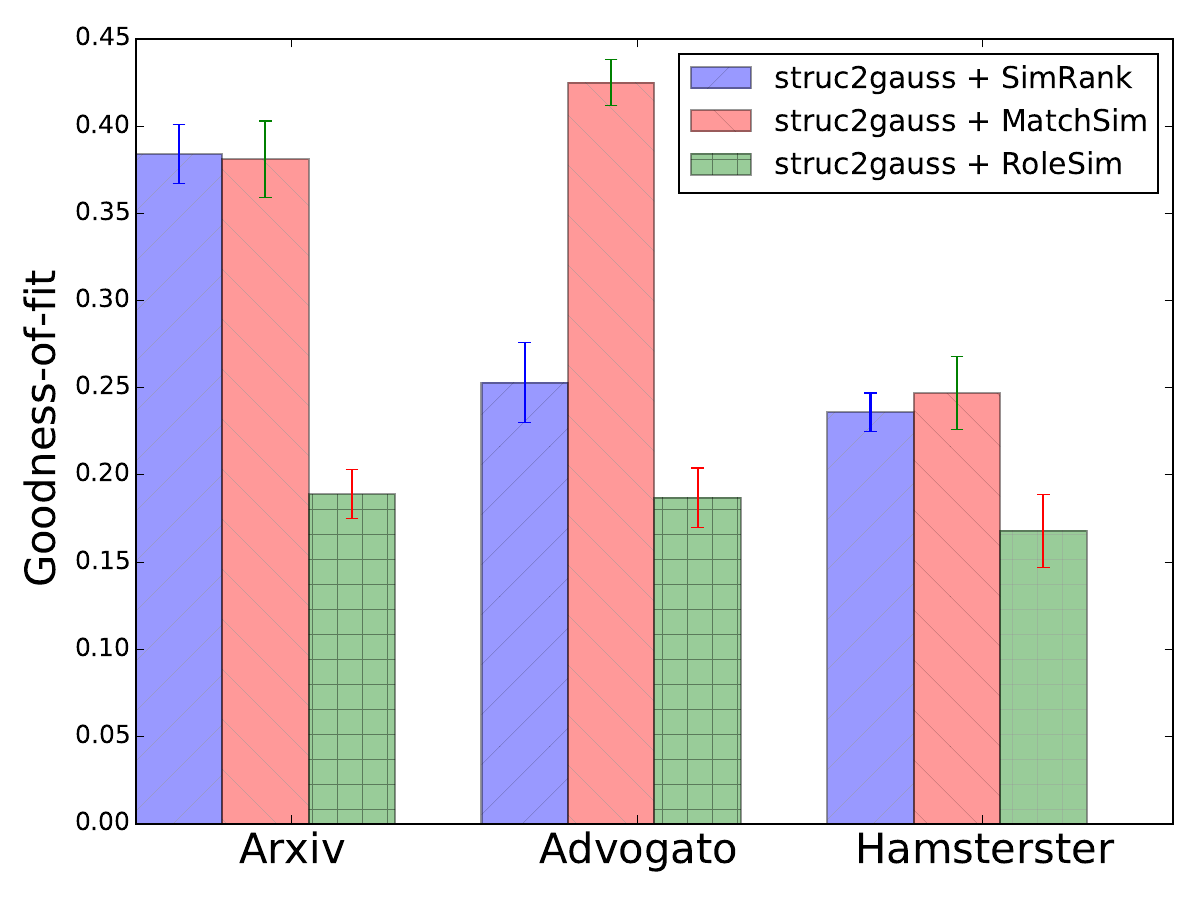}
\caption{Goodness-of-fit of \textit{struc2gauss} with different similarity measures. Lower values are better.}
\label{fig:sim}
\end{figure}

The NMI values for networks with labels are shown in Table~\ref{tb:sim} and the \textit{goodness-of-fit} values are shown in Fig.~\ref{fig:sim}. We can come to the following conclusions:
\begin{itemize}
    \item RoleSim outperforms other two similarity measures in both types of networks with and without clustering labels. It indicates RoleSim can better capture the global structural information. Performance of MatchSim varies on different networks and is similar to \textit{struc2vec}. Thus, it can capture the global structural information to some extent.
    \item SimRank performs worse than other similarity measures as well as \textit{struc2vec} (Table~\ref{tb:air}). Considering the basic assumption of SimRank that "two objects are similar if they relate to similar objects", it computes the similarity also via relations between nodes so that the mechanism is similar to random walk based methods which have been proved not being capable of capturing the global structural information~\citep{lyu2017enhancing}.
\end{itemize}

\subsection{Parameter Sensitivity}
\label{paramstuning}
We consider two types of parameters in \textit{struc2gauss}: (1) parameters also used in other NE methods including \textit{latent dimensions}, \textit{number of samples per node} and \textit{number of positive/negative nodes per node}; and (2) parameters only used in Gaussian embedding including mean constraint $C$ and covariance constraint $c_{max}$ (note that we fix the minimal covariance $c_{min}$ to be 0.5 for simplicity). In order to evaluate how changes to these parameters affect performance, we conducted the same node clustering experiment on the labeled USA-air network introduced in Section~\ref{cluster}. In the interest of brevity, we tune one parameter by fixing all other parameters. In specific, the number of latent dimensions varies from 10 to 200, the number of samples varies from 5 to 15 and the number of positive/negative nodes varies from 40 to 190. Mean constraint $C$ is from 1 to 10, and covariance constraint $c_{max}$ ranges from 1 to 10.

\begin{figure}
\centering
    \begin{subfigure}[b]{0.75\textwidth}
        \centering
        \includegraphics[width=\textwidth]{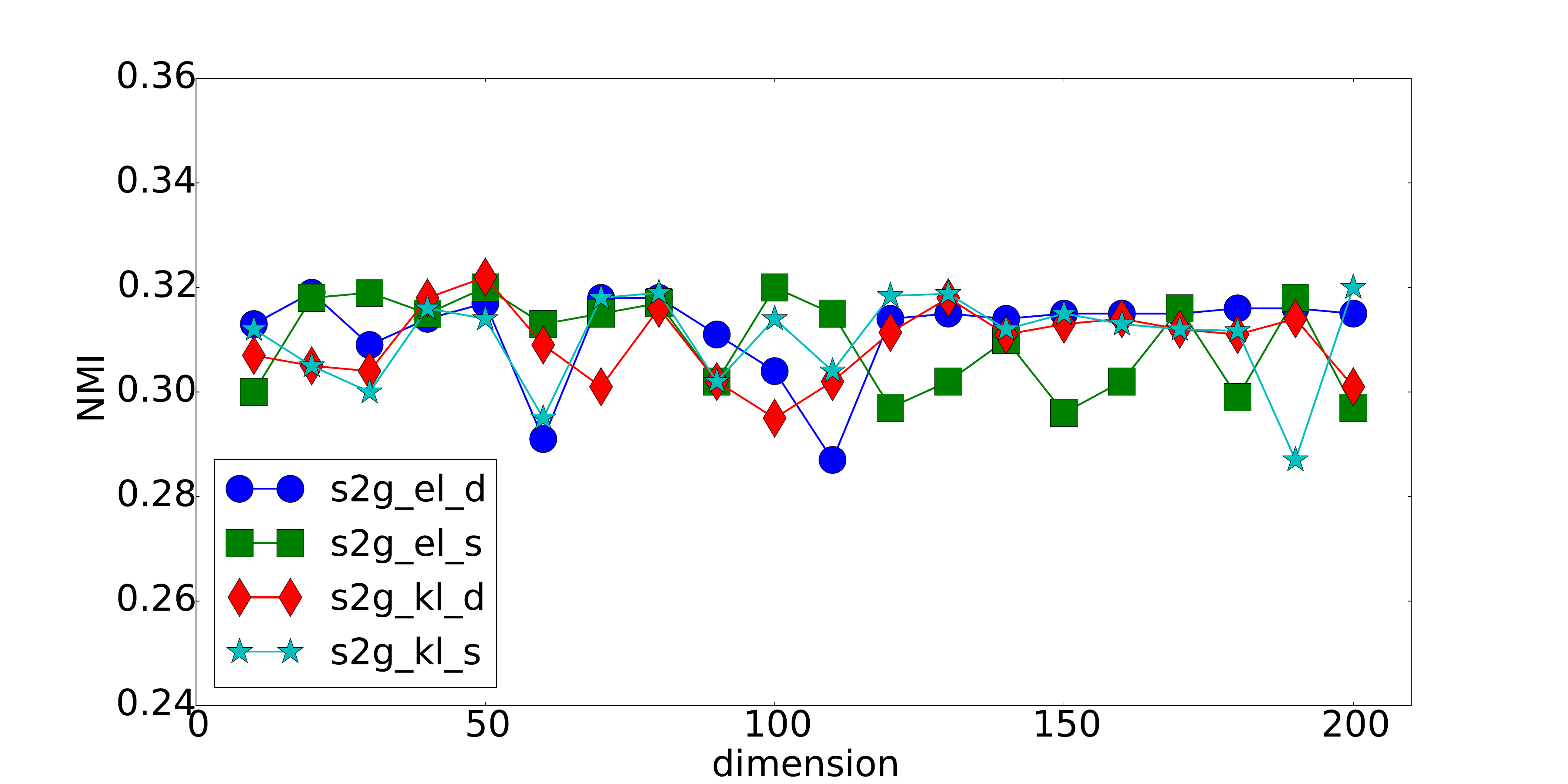}
        \caption[]%
        {{\small Representation dimensions vs. NMI.}}
        \label{fig:dim}
    \end{subfigure}
    \vskip\baselineskip
    \begin{subfigure}[b]{0.75\textwidth}
        \centering
        \includegraphics[width=\textwidth]{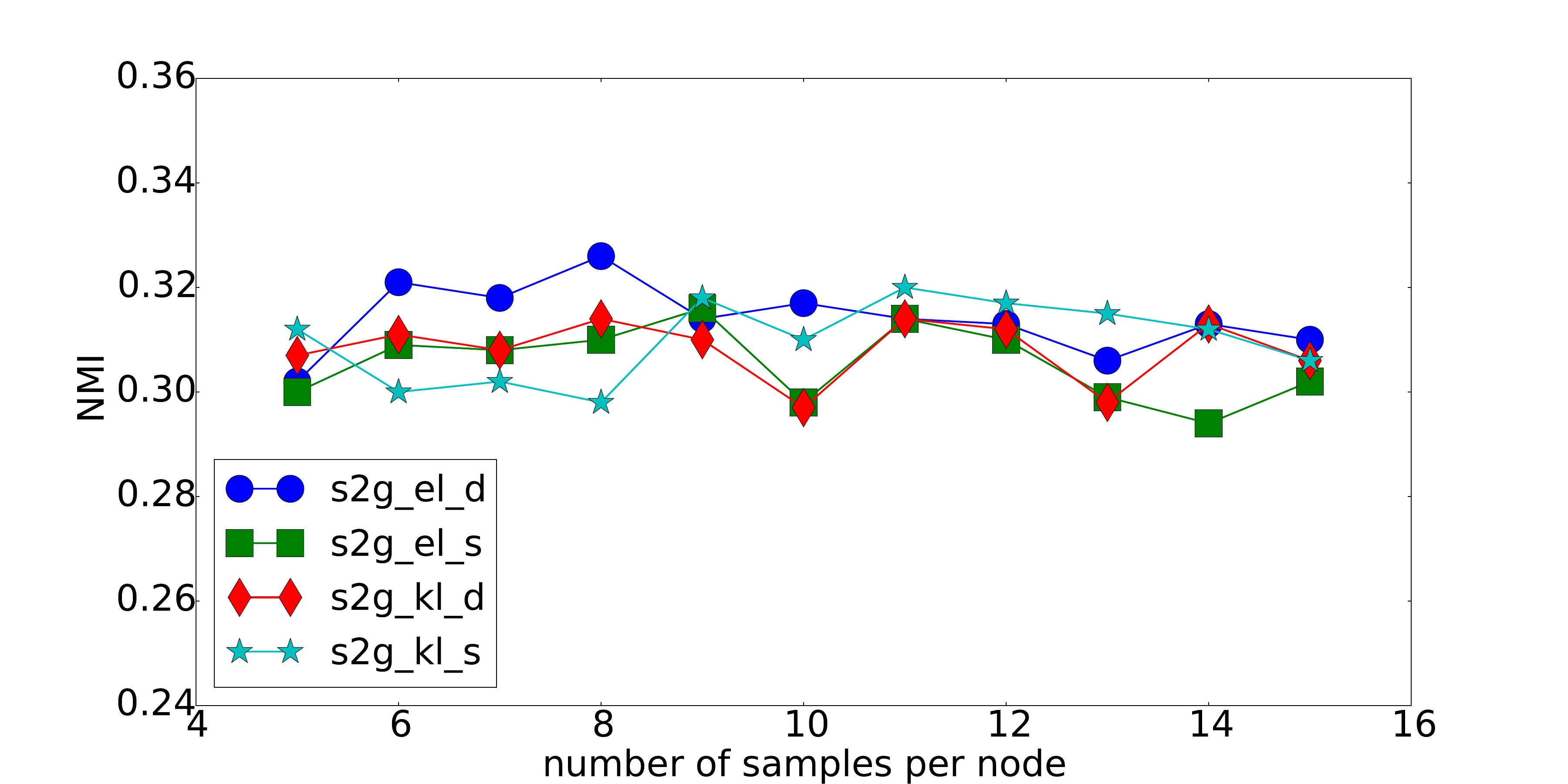}
        \caption[]%
        {{\small Number of samples per node vs. NMI.}}
        \label{fig:sam}
    \end{subfigure}
    \vskip\baselineskip
    \begin{subfigure}[b]{0.75\textwidth}
        \centering
        \includegraphics[width=\textwidth]{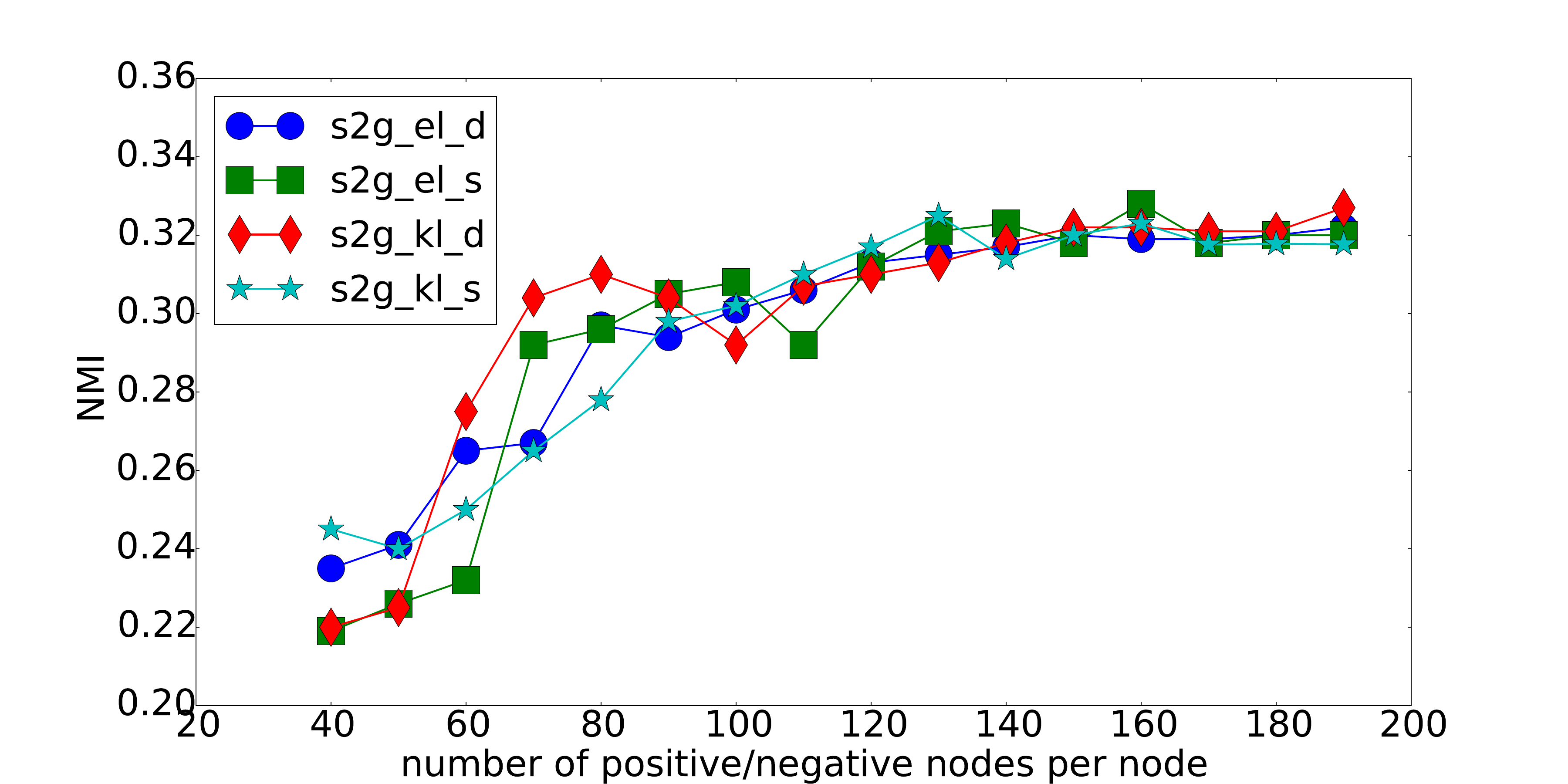}
        \caption[]%
        {{\small Number of positive/negative nodes per node vs. NMI.}}
        \label{fig:walk}
    \end{subfigure}
    \caption[]
    {Parameter Sensitivity Study.}
    \label{fig:param}
\end{figure}

\begin{figure}
\centering
    \begin{subfigure}[b]{0.75\textwidth}
        \centering
        \includegraphics[width=\textwidth]{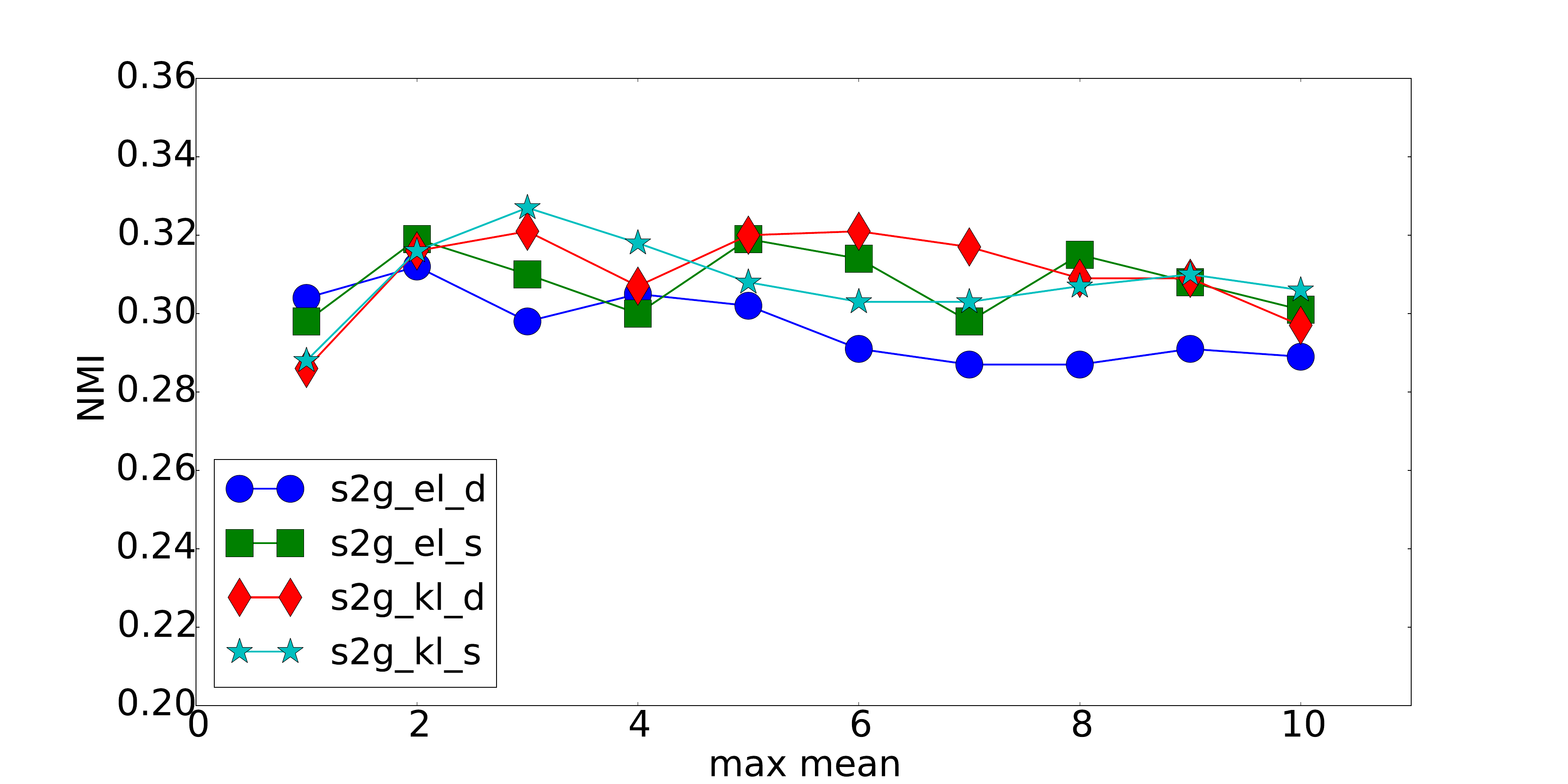}
        \caption[]%
        {{\small mean constraint $C$ vs. NMI.}}
        \label{fig:mu}
    \end{subfigure}
    \vskip\baselineskip
    \begin{subfigure}[b]{0.75\textwidth}
        \centering
        \includegraphics[width=\textwidth]{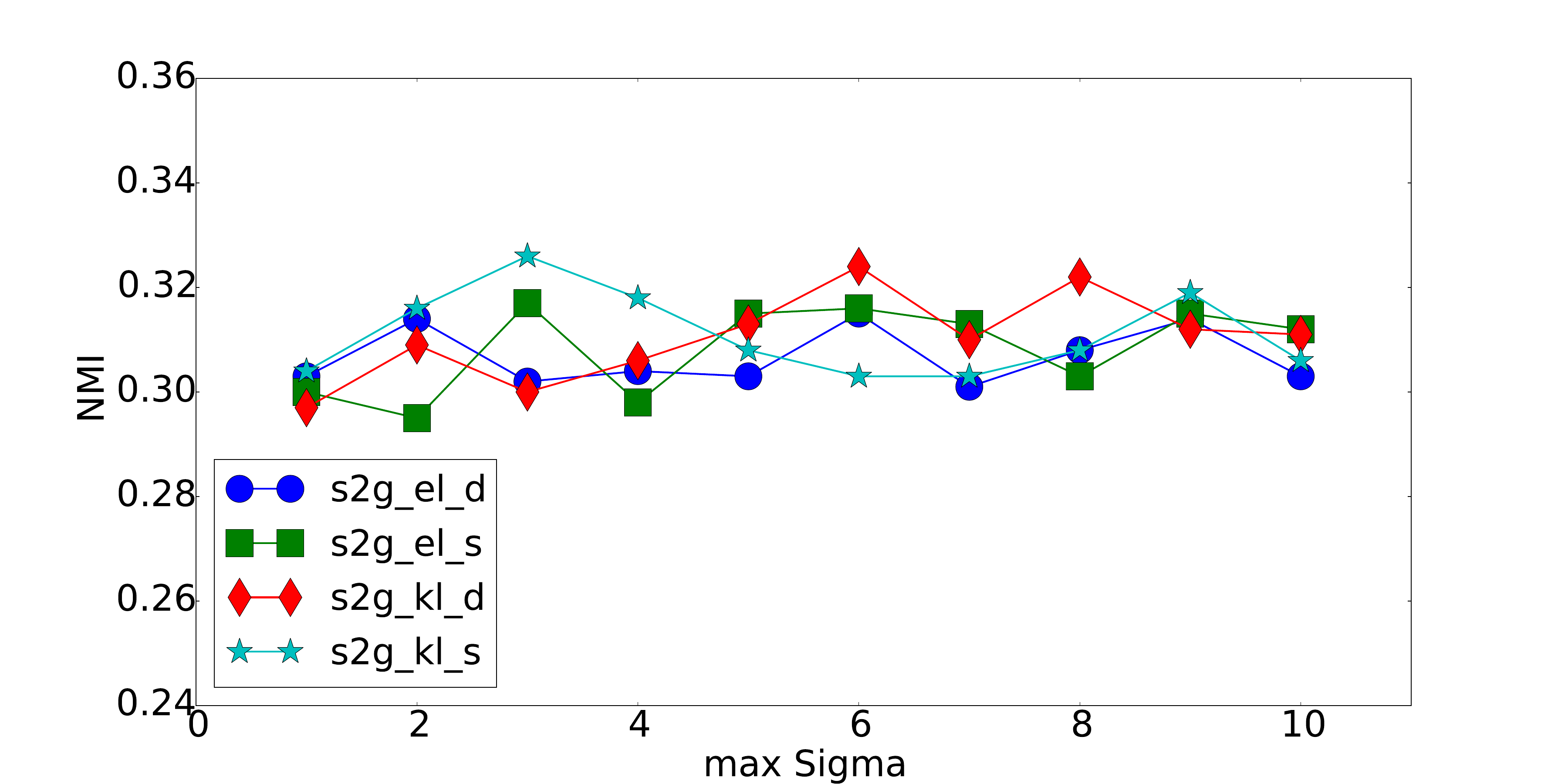}
        \caption[]%
        {{\small covariance constraint $c_{max}$ vs. NMI.}}
        \label{fig:sigma}
    \end{subfigure}
    \caption[]
    {Parameter sensitivity in Gaussian distributions.}
    \label{fig:gauss}
\end{figure}

The results of parameter sensitivity are shown in Fig.~\ref{fig:param} and Fig.~\ref{fig:gauss}. It can be observed from Fig.~\ref{fig:param} (a) and \ref{fig:param} (b) that the trends are relatively stable, i.e., the performance is insensitive to the changes of representation dimensions and numbers of samples. The performance of clustering is improved with the increase of numbers of positive/negative nodes shown in Fig.~\ref{fig:param} (c). Therefore, we can conclude that \textit{struc2guass} is more stable than other methods. It has been reported that other methods, e.g., DeepWalk~\citep{perozzi2014deepwalk}, LINE~\citep{tang2015line} and \textit{node2vec}~\citep{grover2016node2vec}, are sensitive to many parameters. In general, more dimensions, more walks and more context can achieve better performance. However, it is difficult to search for the best combination of parameters in practice and it may also lead to overfitting. For Gaussian embedding specific parameters $C$ and $c_{max}$, both trends are stable, i.e., the selection of these contraints have little effect on the performance. Although with larger mean constraint $C$, the NMI decreases but the difference is not huge.

\begin{table}
\small
\centering
\caption{NMI for node clustering in Brazil-air network with different numbers of noisy edges.}
\label{tb:noise1}
\begin{tabular}{|l|c|c|c|c|c|c|c|}
\hline
\# noisy edges     & 0     & 50     & 100    & 150   & 200   & 250  &  300 \\ \hline
\textit{graph2gauss}  & 0.1204  & 0.1032   & 0.0903  &  0.0913 & 0.0852 & 0.0833 &  0.0683     \\ \hline
$s2g$\_el\_d  & 0.5615  & 0.5165   & 0.5161  &  0.5122 & 0.4810 & 0.4754 &  0.4787     \\ \hline
$s2g$\_el\_s  & 0.5396  & 0.4338   & 0.4180  &  0.4152 & 0.4102 & 0.3956 &  0.3924     \\ \hline
$s2g$\_kl\_d  & 0.5527  & 0.5186   & 0.5036  &  0.4940 & 0.4824 & 0.4736 &  0.4103     \\ \hline
$s2g$\_kl\_s  & 0.5527  & 0.5310   & 0.5214  &  0.4951 & 0.4895 & 0.4621 &  0.4651     \\ \hline
\end{tabular}
\end{table}
\begin{table}
\small
\centering
\caption{NMI for node clustering in Europe-air network with different numbers of noisy edges.}
\label{tb:noise2}
\begin{tabular}{|l|c|c|c|c|c|c|c|}
\hline
\# noisy edges     & 0     & 500     & 1000    & 1500   & 2000   & 2500  &  3000 \\ \hline
\textit{graph2gauss}  & 0.1109  & 0.0776   & 0.0727  &  0.0716 & 0.0634 & 0.0702 &  0.0613     \\ \hline
$s2g$\_el\_d  & 0.3234  & 0.1767   & 0.1634  &  0.1694 & 0.1492 & 0.1431 &  0.1413     \\ \hline
$s2g$\_el\_s  & 0.2974  & 0.1613   & 0.1505  &  0.1432 & 0.1452 & 0.1285 &  0.1042     \\ \hline
$s2g$\_kl\_d  & 0.3145  & 0.2664   & 0.2014  &  0.1854 & 0.1802 & 0.1634 &  0.1361     \\ \hline
$s2g$\_kl\_s  & 0.3280  & 0.3024   & 0.2930  &  0.1504 & 0.1514 & 0.1414 &  0.1367     \\ \hline
\end{tabular}
\end{table}

\subsection{Efficiency and Effectiveness Study}
As discussed above in Section~\ref{complex}, the high computational complexity is one of the major issues in our method. In this experiment, we empirically study this computational issue by comparing the run-time and performance of different global structural preserving baselines and a heuristic method to accelerate the RoleSim measures. The heuristic method, named Fast \textit{struc2gauss}, is introduced in Section~\ref{complex}: we set the similarity to be 0 if two nodes have a large difference in degrees to avoid more computing for dissimilar node pairs. For simplicity, we only test \textit{struc2gauss} with KL and spherical covariance. Also, we only consider embedding methods that can preserve the structural role information as baselines, i.e., GraphWave, \textit{struc2vec} and DRNE.

We conduct the experiments on the larger networks without ground-truth labels because on smaller networks the run-time differences are not significant. The run-time comparison is shown in Table~\ref{tb:time} and the performance comparison is shown in Table~\ref{tb:perfm}. Note that NA in these tables because these methods reported a memory error and did not obtain any results. To make a fair comparison, all these methods are run in the same machine with 128GB memory and GPU have not been used for DRNE. From these results, it can be observed: (1) although the computational issue still exists, our method can achieve good performance compared to state-of-the-art structural role preserving network embedding methods such as GraphWAVE and \textit{struc2vec}. (2) Although DRNE is much fast, its performance is worse than our method and other baselines. Moreover, it is incapable of modeling uncertainties. (3) Fast \textit{struc2gauss} can effectively accelerate RoleSim computing and achieve comparable performance in role clustering.
\begin{table}
\small
\centering
\caption{Run-time for different structural role preserving network embedding methods.}
\label{tb:time}
\begin{tabular}{|l|c|c|c|c|c|}
\hline
         & GraphWAVE     & \textit{struc2vec}    &  DRNE    &   \textit{struc2gauss}  &   Fast \textit{struc2gauss}     \\ \hline
Arxiv & 90.68s         & 10+h         & 159.43s  &  2h  &  886.93s\\ \hline
Advogato  & 172.13s         & 10+h         & 191.52s   & 4h  & 1962.68s \\ \hline
Hamsterster  & 24.25s         & 10+h         & 85.93s  &  1h  &  456.24s \\ \hline
anybeat  & NA         & NA         & 1094.64s   & 13h  & 5h \\ \hline
Epinion  & NA         & NA         & 2938.83s   & 20h+  & 12h \\ \hline
\end{tabular}
\end{table}
\begin{table}
\small
\centering
\caption{Performance (\textit{goodness-of-fit}) of different structural role preserving network embedding methods.}
\label{tb:perfm}
\begin{tabular}{|l|c|c|c|c|c|}
\hline
         & GraphWAVE     & \textit{struc2vec}    &  DRNE    &   \textit{struc2gauss}  &   Fast \textit{struc2gauss}     \\ \hline
Arxiv & 0.5435         & 0.3674         & 0.6822  &  0.1880  &  0.1983\\ \hline
Advogato  & 0.3938         &  0.2751         & 0.6102   & 0.1852  & 0.2012 \\ \hline
Hamsterster  & 0.3385         & 0.1878         & 0.5939  &  0.1666  &  0.1790 \\ \hline
anybeat  & NA         & NA         & 0.5639   & 0.1597  & 0.1622 \\ \hline
Epinion  & NA         & NA         & 0.4978   & 0.2270  & 0.2452 \\ \hline
\end{tabular}
\end{table}

\section{Discussion}
\label{dis}
The proposed \textit{struc2gauss} is a flexible framework for node representations. As shown in Fig.~\ref{fig:frame}, different similarity measures can be incorporated into this framework and empirical studies will be presented in Section~\ref{sim}. Furthermore, other types of methods which model structural information can be utilized in \textit{struc2gauss} as well.

To illustrate the potential to incorporate different methods, we categorize different methods for capturing structural information into three types:
\begin{itemize}
    \item \textbf{Similarity-based methods}. Similarity-based methods calculate pairwise similarity based on the structural information of a given network. Related work has been reviewed in Section~\ref{strucsim}.
    \item \textbf{Ranking-based methods}. PageRank~\citep{page1999pagerank} and HITS~\citep{kleinberg1999authoritative} are two most representative ranking-based methods which learns the structural information. PageRank has been used for NE in~\citep{ma2017preserving}. 
    \item \textbf{Partition-based methods}. This type of methods, e.g., role discovery, aims to partition nodes into disjoint or overlapping groups, e.g., REGE~\citep{borgatti1993two} and RolX~\citep{henderson2012rolx}.
\end{itemize}
In this paper, we focus on \textbf{similarity-based methods}. For \textbf{ranking-based methods}, we can use a fixed sliding window on the ranking list, then given a node the nodes within the window can be viewed as the context. In fact, this mechanism is similar to DeepWalk. For \textbf{partition-based methods}, we can consider the nodes in the same group as the context for each other.

\section{Conclusions and Future Work}
\label{conc}
Two major limitations exist in previous NE studies: i.e., \textbf{structure preservation} and \textbf{uncertainty modeling}. Random-walk based NE methods fail in capturing global structural information and representing a node into a point vector are not capable of modeling the uncertainties of node representations.

We proposed a flexible structure preserving network embedding framework, \textit{struc2gauss}, to tackle these limitations. On the one hand, \textit{struc2gauss} learns node representations based on structural similarity measures so that global structural information can be taken into consideration. On the other hand, \textit{struc2gauss} utilizes Gaussian embedding to represent each node as a Gaussian distribution where the mean indicates the position of this node in the embedding space and the covariance represents its uncertainty.

We experimentally compared three different structural similarity measures for networks and two different energy functions for Gaussian embedding. By conducting experiments from different perspectives, we demonstrated that \textit{struc2gauss} excels in capturing global structural
information, compared to state-of-the-art NE techniques such as DeepWalk, \textit{node2vec} and \textit{struc2vec}. It outperforms other competitor methods in role discovery task and structural role classification on several real-world networks. It also overcomes the limitation of uncertainty modeling and is capable of capturing different levels of uncertainties. Additionally, \textit{struc2gauss} is less sensitive to different parameters which makes it more stable in practice without putting more effort in tuning parameters.

In the future, we will explore faster RoleSim measures for more scalable NE methods, for example, fast method to select $k$ most similar nodes for a given node. Also, it is a promising research direction to investigate different strategies to model global structural information except structural similarity in NE tasks. Besides, other future investigations in this area include learning node representations in dynamic and temporal networks.


\bibliographystyle{spbasic}      
\bibliography{ref.bib}

\end{document}